\documentclass[letterpaper,prl,twocolumn,amsmath,amssymb,superscriptaddress,longbibliography]{revtex4-1}
\usepackage[utf8]{inputenc}
\usepackage[draft]{hyperref}
\usepackage{graphicx}
\usepackage{dcolumn}
\usepackage{bm}
\usepackage{amssymb}
\usepackage{physics}
\usepackage{color}
\usepackage{soul}
\usepackage{epstopdf}
\usepackage{float}

\begin{document}
\title{Complete polarization control for a nanofiber waveguide using directional coupling}

\author{Fuchuan Lei}
\author{Georgiy Tkachenko}
\email[Corresponding author: ]{georgiy.tkachenko@oist.jp}
\thanks{the first two authors contributed equally to this work.}
\author{Jonathan M. Ward}
\author{S\'{i}le Nic Chormaic}
\altaffiliation[Also at ]{Institut N\'eel, Universit\'e Grenoble Alpes, F-38042 Grenoble, France}
\affiliation{Light-Matter Interactions Unit, Okinawa Institute of Science and Technology Graduate University, Onna, Okinawa 904-0495, Japan}

\date{\today}

\begin{abstract}
Optical nanofiber waveguides are widely used for near-field delivery and measurement of light. Despite their versatility and efficiency, nanofibers have a critical drawback - their inability to maintain light's polarization state on propagation. Here, we design a directional coupler consisting of two crossed nanofibers  to probe the polarization state at the waist region. Directionality of coupling occurs due to asymmetric dipolar emission or spin-locking when the evanescent field pattern breaks the mirror symmetry of the crossed-nanofiber system. We demonstrate that, by monitoring the outputs from the  directional coupler, two non-orthogonal polarization states can be prepared at the nanofiber waist with a fidelity higher than 99\%. Based on these states, we devise a simple and reliable method for complete control of the polarization along a nanofiber waveguide.
\end{abstract}

\maketitle

\section{I. Introduction}
\begin{figure}
\centering
\includegraphics[width=0.85\linewidth]{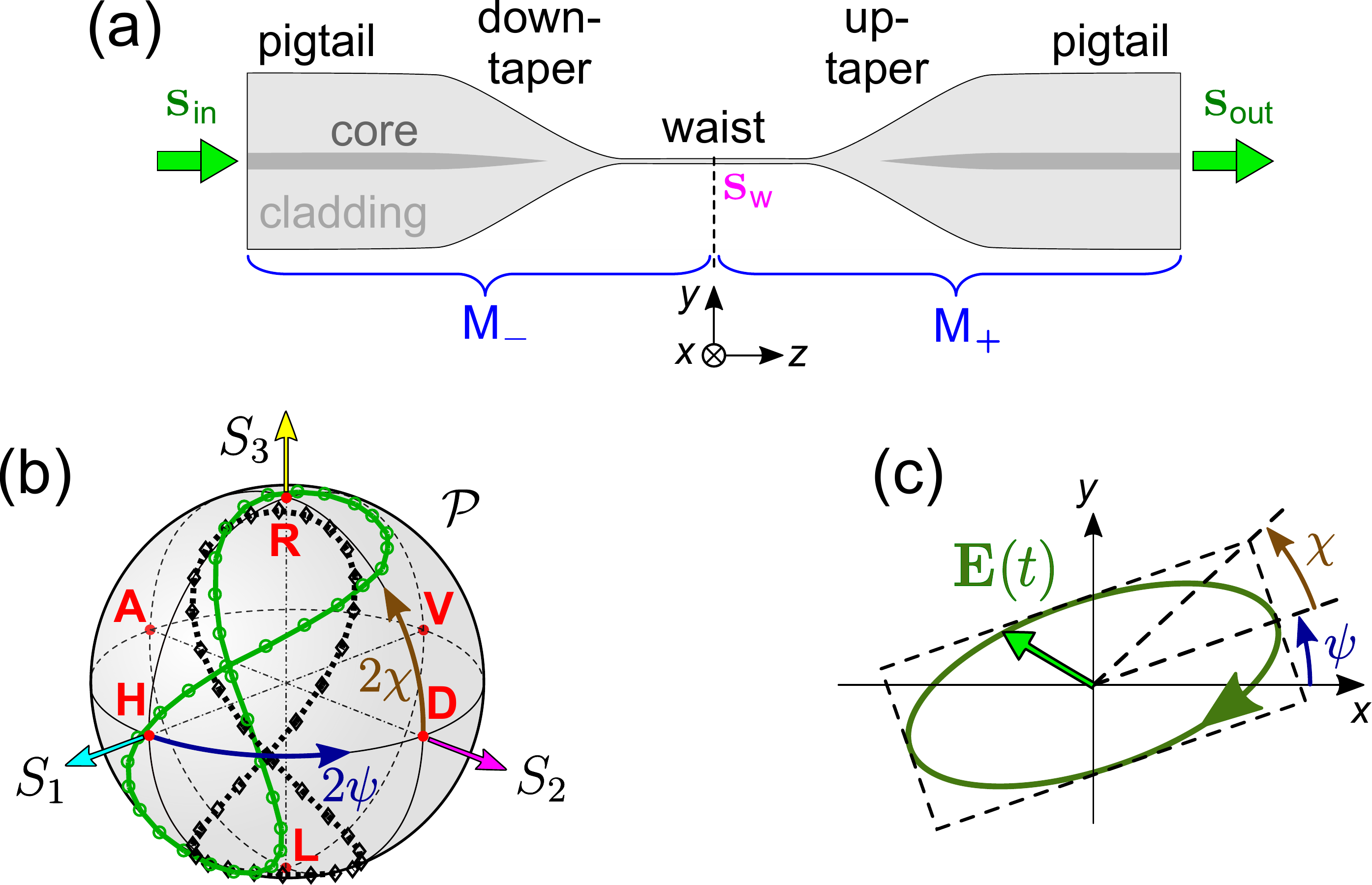}
\caption{(a)~A tapered optical fiber with an ultrathin waist does not, generally, maintain the polarization of guided light. Transformation of the input polarization state, ${\bf s}_{\rm in}$, to that at the waist, ${\bf s}_{\rm w}$, is described by the Mueller matrix ${\rm M}_-$. Another matrix, ${\rm M}_+\neq{\rm M}_-$, describes the transformation into the output state, ${\bf s}_{\rm out}$. (b)~The trajectories traced on the Poincar\'e sphere, $\cal P$, by the input (${\bf s}_{\rm in}$, diamonds and dotted line) and the output (${\bf s}_{\rm out}$, circles and solid line) polarization states are different, even for short and straight nanofibers. (c)~Polarization ellipse defined by the angles $\psi$ and $\chi$, which correspond to half the azimuthal and polar angles on $\cal P$.}
\label{fig:nanofiber}
\end{figure}

Tapered optical fibers are unique because they allow for the smooth transition of light from macro- to micro- or nanoscale systems. Owing to the strong evanescent field around the ultrathin waist region, such fibers are efficient and versatile tools for optical manipulation~\cite{brambilla_ol_2007,maimaiti_sr_2015,ren2016tapered}, sensing~\cite{yoshie_sensors_2011,yu_am_2014}, nonlinear optics~\cite{beugnot2014brillouin,zoubi_prl_2017}, microcavities~\cite{cai_prl_2000}, atomic physics~\cite{le2004atom,hendrickson_prl_2010,vetsch_ieee_2012,kumar_njp_2015,kornovan_prb_2018}, and various studies of light-matter interactions  in both classical and quantum regimes~\cite{tong_nl_2009,tong_oc_2012,solano_chapter_2017}.
As  light propagates from the fiber pigtail to the waist region (Fig.~\ref{fig:nanofiber}(a)), the mode volume reduces, and the intensity of the evanescent field increases~\cite{le_kien_oc_2004}. Optical power can be transferred to the waist without significant losses, even for ultrathin fibers. However, the other important parameter of the field---its polarization---is usually impossible to control.  The polarization of the evanescent field is crucial for many nanofiber-based studies including microcavity mode excitation~\cite{knight_ol_1997} and  interactions with nanoparticles~\cite{wang_oc_2007} or atomic ensembles~\cite{nieddu_jo_2016, sadgrove_sr_2016}.

As simple as it may seem, the problem of polarization uncertainty in tapered fibers has challenged the scientific community for decades. To solve it, one must establish a direct link between the near field at the ultrathin waist and the far field where macroscopic detectors and controllers are typically placed. In this work, we explore directional coupling between two crossed nanofibers, and, consequently, develop a reliable method for achieving an arbitrary polarization state in the evanescent field of a single-mode nanofiber.
The physics behind this directionality features the interplay of two phenomena: the well-known spin-momentum locking, and the newly found mirror-symmetry breaking of a light-matter system by an induced electric dipole.

We begin by introducing optical nanofibers and their fabrication in Sec.~II, and discuss how the polarization of light propagates in adiabatically tapered fibers. We emphasize that the polarization transformation in such fibers, as in any optical elements free of depolarization and dichroism, is equivalent to rotations of the Poincar\'e sphere.
In Sec.~III, we devise a simple, two-step procedure, which allows for reversing the above rotational transformations via consecutive mapping of two non-orthogonal polarization states. By compensating an arbitrary transformation, one achieves  complete control over the polarization state.
Section~IV is dedicated to the directional coupling between two single-mode optical nanofibers crossed at right angles. We present a detailed experimental and numerical study of the directional coupler's operation. The results indicate how to securely identify a pair of non-orthogonal states required for the two-step compensation procedure to be applicable in the case of a nanofiber waveguide.
Section~V presents a practical demonstration of the polarization control. We discuss its precision and accuracy, as well as the experimental evidence related to polarization evolution in a tapered optical fiber.

\section{II. Polarization of light in a nanofiber waveguide}
Nanofibers are generally produced from conventional optical fibers by controlled heating and pulling \citep{ward_rsi_2014}. A typical, single-mode nanofiber consists of a cylindrical submicron-diameter waist connected to  fiber pigtails by two taper regions, see Fig.~\ref{fig:nanofiber}(a), where the Cartesian coordinate system $(x,y,z)$ originates in the middle of the waist and $z$ is parallel to the fiber axis. For this work, the tapered fibers were prepared from a step-index cylindrical optical fiber with a cut-off wavelength of $920\pm50$~nm. The cylindrical waist regions were about 2~mm long with radii of $159\pm3$~nm, as measured by a scanning electron microscope. The input pigtail was coupled to a collimated Gaussian beam from a continuous wave laser (980~nm wavelength). Each fiber was kept as short, straight, and strain-free as possible. Such precautions are common practice in nanofiber experiments where polarization (or mode) transformations are undesirable.

The key questions to be addressed are (i) how  does the polarization of  guided light change upon propagation through a tapered fiber? and---more importantly---(ii) how can this change be controlled? A polarization state can be treated as a unit vector ${\bf s}=(1,S_1,S_2,S_3)$, where $S_{1,2,3}$ are the Stokes parameters which define a point on the Poincar\'e sphere, $\cal P$, see Fig.~\ref{fig:nanofiber}(b). The azimuthal ($2\psi$) and polar ($2\chi$) angles on the sphere are directly linked to the orientation, ellipticity, and handedness (through the sign of $\chi$) of the polarization ellipse traced in the $(x,y)$ plane by the tip of the electric field vector, ${\bf E}(t)$, see Fig.~\ref{fig:nanofiber}(c), in the following way: ${\bf s}=(1,\cos2\psi\cos2\chi,\sin2\psi\cos2\chi,\sin2\chi)$.

First, we compared the polarization state at the input, ${\bf s}_{\rm in}$, and the output, ${\bf s}_{\rm out}$, of our tapered fibers. In order to probe a large part of $\cal P$, ${\bf s}_{\rm in}$ was driven around the Poincar\'e sphere in a figure-of-eight trajectory by passing the horizontally-polarized ($\bf H$) input beam through a rotating quarter-wave plate (QWP). The trajectory was recorded by means of a commercial polarization analyzer. For every fiber under test, the output trajectory repeated the shape of the input, but always exhibited a significant shift on the sphere (see a typical example in Fig.~\ref{fig:nanofiber}(b)), regardless of our efforts to maintain the polarization using the fore-mentioned precautions.

To understand how the shape of the trajectory on $\cal P$ is maintained, let us note that the fibers were designed according to the adiabatic condition which implies energy transfer with minimum loss.  Adiabaticity was ensured by keeping the taper angle below the critical value~\cite{love_ieee_1991, jung_oe_2008} and the transmission at 980~nm wavelength above 97\% throughout the pulling process. Both pigtails and the cylindrical waist region are single-mode, hence losses can only occur in the tapers via coupling to radiation modes and higher-order modes which cannot propagate along the single-mode fiber. The guided electric field, ${\bf E}(z)$, in adiabatic single-mode fibers can be described as a combination of two orthogonal, quasi-linearly polarized fundamental eigenmodes, ${\bf HE}_{11}^{x}$ and ${\bf HE}_{11}^{y}$~\cite{le_kien_oc_2004}:
\begin{equation}
{\bf E}(z)=\alpha(z)\,{\bf HE}_{11}^{x}+\beta(z)\,{\bf HE}_{11}^{y}\,,
\end{equation}
where $\alpha$ and $\beta$ are variable complex amplitudes. The polarization ellipse is associated with the Jones vector
\begin{equation}
{\bf j}=\frac{1}{\sqrt{|\alpha|^2+|\beta|^2}}\,\begin{bmatrix}
\alpha\\
\beta
\end{bmatrix}\,.
\end{equation}
Its evolution upon propagation from $z_0$ to $z$ through the fiber can be written as
\begin{equation}
{\bf j}(z)=u_{\rm fiber}\,{\bf j}(z_0)\,,
\end{equation}
 and, given that adiabatic fibres have transmission close to unity, $u_{\rm fibre}$ must be a $2\times2$ unitary matrix: 
\begin{equation}
u_{\rm fibre}^{\dagger}u_{\rm fibre}={\rm I}\,,
\label{eq:unitary}
\end{equation}
where $\rm I$ is the identity matrix.
Therefore, transformations of the Stokes vectors in adiabatically-tapered fibers are restricted to the 3D rotation (SO(3)) group~\cite{sakurai1994quantum}. Rotations of the Poincar\'e sphere (${\cal P}\to{\cal P}'$) preserve angles between the Stokes vectors, and, consequently, the shape of the trajectories for ${\bf s}_{\rm in}$ and ${\bf s}_{\rm out}$ in Fig.~\ref{fig:nanofiber}(b). This preservation of angles is the key condition for the polarization control that we achieve.

\section{III. Two-step polarization compensation}
The polarization state at the nanofiber waist, ${\bf s}_{\rm w}$, can be linked to the input and output states in terms of Mueller calculus~\cite{optics_handbook}: ${\bf s}_{\rm w}={\rm M}_-{\bf s}_{\rm in}$, and ${\bf s}_{\rm out}={\rm M}_+{\bf s}_{\rm w}$, where ${\rm M}_-$ and ${\rm M}_+$ are $4\times4$ matrices describing the Stokes vector evolution before ($z<0$) and after ($z>0$) the waist. These two matrices do not correlate and  cannot be determined if one only measures ${\bf s}_{\rm in}$ and ${\bf s}_{\rm out}$. Consequently, in order to control ${\bf s}_{\rm w}$, one has to probe the evanescent field.

\begin{figure}
\centering
\includegraphics[width=0.75\linewidth]{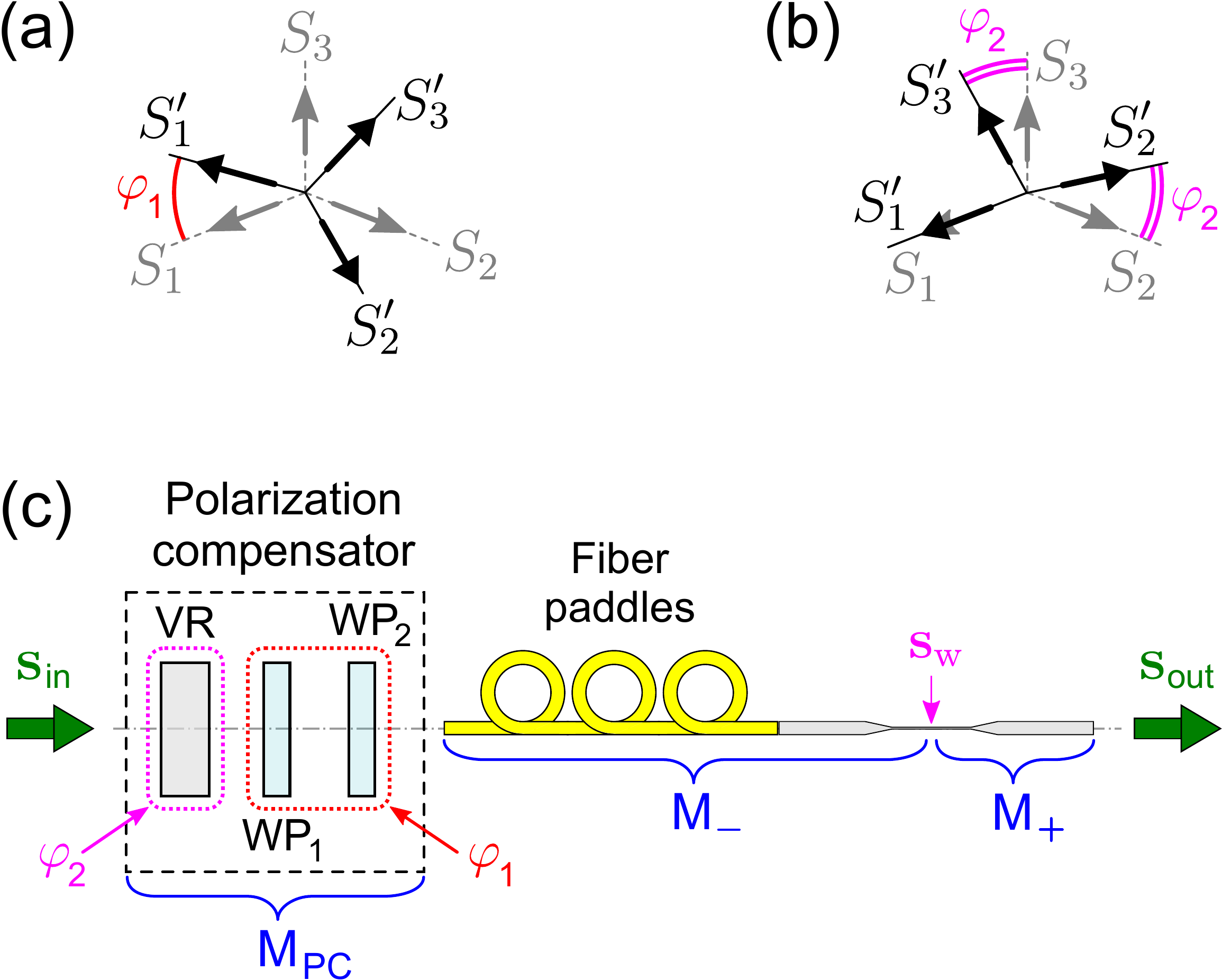}
\caption{(a),(b)~An unitary transformation of the Poincar\'e sphere $\cal P \to \cal P'$ can be decomposed into two independent rotations by angles $\varphi_1$ and $\varphi_2$. To reverse the transformation, we adjust $\varphi_1$ in order to achieve one state (here $\bf H$ or $S_1=1$) maintained (that is ${\bf H} \to {\bf H}$, $S_1'=S_1=1$). Then, while keeping $\varphi_1$ fixed, $\varphi_2$ is adjusted until $S_{2,3}'=S_{2,3}$. (c)~To control polarization at the nanofiber waist, we compensate for the unknown matrix ${\rm M}_-$ by sequentially picking $\varphi_1$ and $\varphi_2$ using two quarter-wave plates (${\rm WP}_{1,2}$) and a variable retarder~(VR). Fiber paddles allow randomization of ${\rm M}_-$ when studying the precision of the control.}
\label{fig:concept}
\end{figure}

Theoretically, ${\rm M}_-$ can be found by measuring several sets of (${\bf s}_{\rm in}$,~${\bf s}_{\rm w}$)~\cite{optics_handbook}. However, this procedure is not always realistic. Instead of controlling the target polarization state by deducing ${\rm M}_-$, in this work we followed a different strategy, which is more practical and, as will become clear, is the only option for nanofiber waveguides. Namely, we reverse the transformation of the Poincar\'e sphere, ${\cal P}\to{\cal P}'$, thus achieving ${\cal P}\to{\cal P}$ and ${\bf s}_{\rm w}={\bf s}_{\rm in}$.

From a geometrical point of view, the orientation of a sphere can be completely defined by two independent angles, i.~e., latitude and longitude. It is, therefore, logical  to realize the ${\cal P}\to{\cal P}$ mapping in two steps:
\begin{enumerate}
\item tilting of one axis by an angle $\varphi_1$ (Fig.~\ref{fig:concept}(a));
\item rolling of the two other axes about the first one by an angle $\varphi_2$ (Fig.~\ref{fig:concept}(b)).
\end{enumerate}

Importantly, this two-step procedure for reversing unknown polarization transformations is not restricted to nanofibers, but can be applied to any optical element free of depolarization and dichroism.
In practice, we realized the procedure by means of a free-space polarization compensator (PC) consisting of a variable retarder (VR, with the fast axis parallel to $x$) and a pair of quarter-wave plates (${\rm WP}_{1,2}$), see Fig.~\ref{fig:concept}(c). The compensator is characterized by an unitary Jones matrix, $u_{\rm PC}$, or a Mueller matrix, ${\rm M}_{\rm PC}$. In step~(1), ${\rm WP}_1$ and ${\rm WP}_2$ are independently rotated  until the input horizontal polarization (${\bf s}_{\rm in}={\bf H}$; $S_1=1$) is mapped onto itself at the nanofiber waist (${\bf s}_{\rm w}={\bf H}$; $S_1'=1$). Next, in step~(2), an input state with $|S_1|\neq1$ is selected, and the retardance of VR is adjusted. This drives ${\bf s}_{\rm w}$ along the circle in the plane parallel to $(S_2,S_3)$ until eventually $S_{2,3}'=S_{2,3}$, and thus ${\bf s}_{\rm w}={\bf s}_{\rm in}$ due to ${\rm M}_{\rm PC}={\rm M}_-^{-1}$. In fact, ${\cal P}\to{\cal P}$ mapping can be performed with any pair of {\it non-orthogonal} states, i.~e. such states that do not lie on the same diameter of the Poincar\'e sphere, or---in mathematical terms---have a non-zero inner product (see the proof in Appendix).

\section{IV. Crossed-nanofiber directional coupler}
\begin{figure}
\centering
\includegraphics[width=1\linewidth]{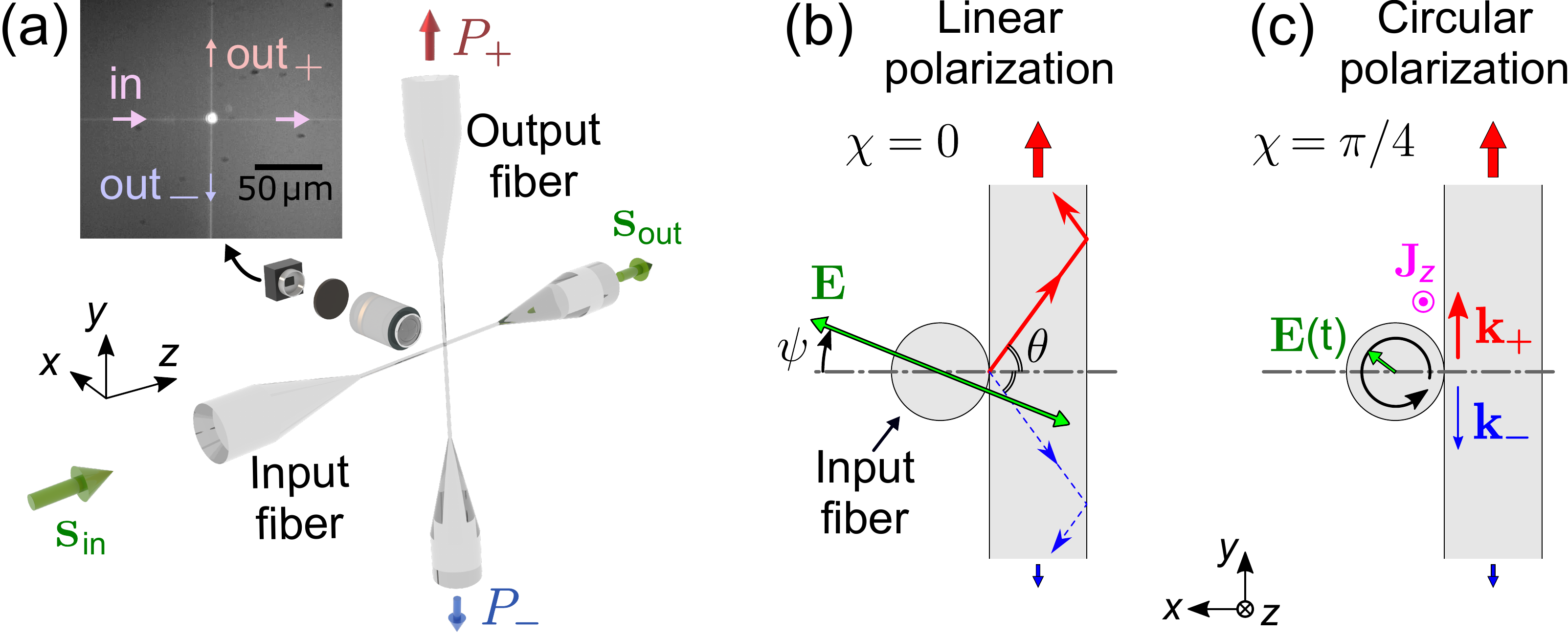}
\caption{(a)~Crossed-nanofiber optical coupler and the imaging system consisting of a $\times20$ objective lens, a linear polarizer, and a video camera. The camera image (top-left) shows nanofibers and the scattering spot from the crossing. (b)~Tilted  linear polarization states of guided light break the mirror symmetry of the system, thus leading to power imbalance between the output channels. (c)~Circular polarization breaks the symmetry too. Directional coupling occurs due to locking of the transverse spin angular momentum, ${\bf J}_z$, to the output wave vector, here ${\bf k}_+$.}
\label{fig:coupler}
\end{figure}

\begin{figure}
\centering
\includegraphics[width=1\linewidth]{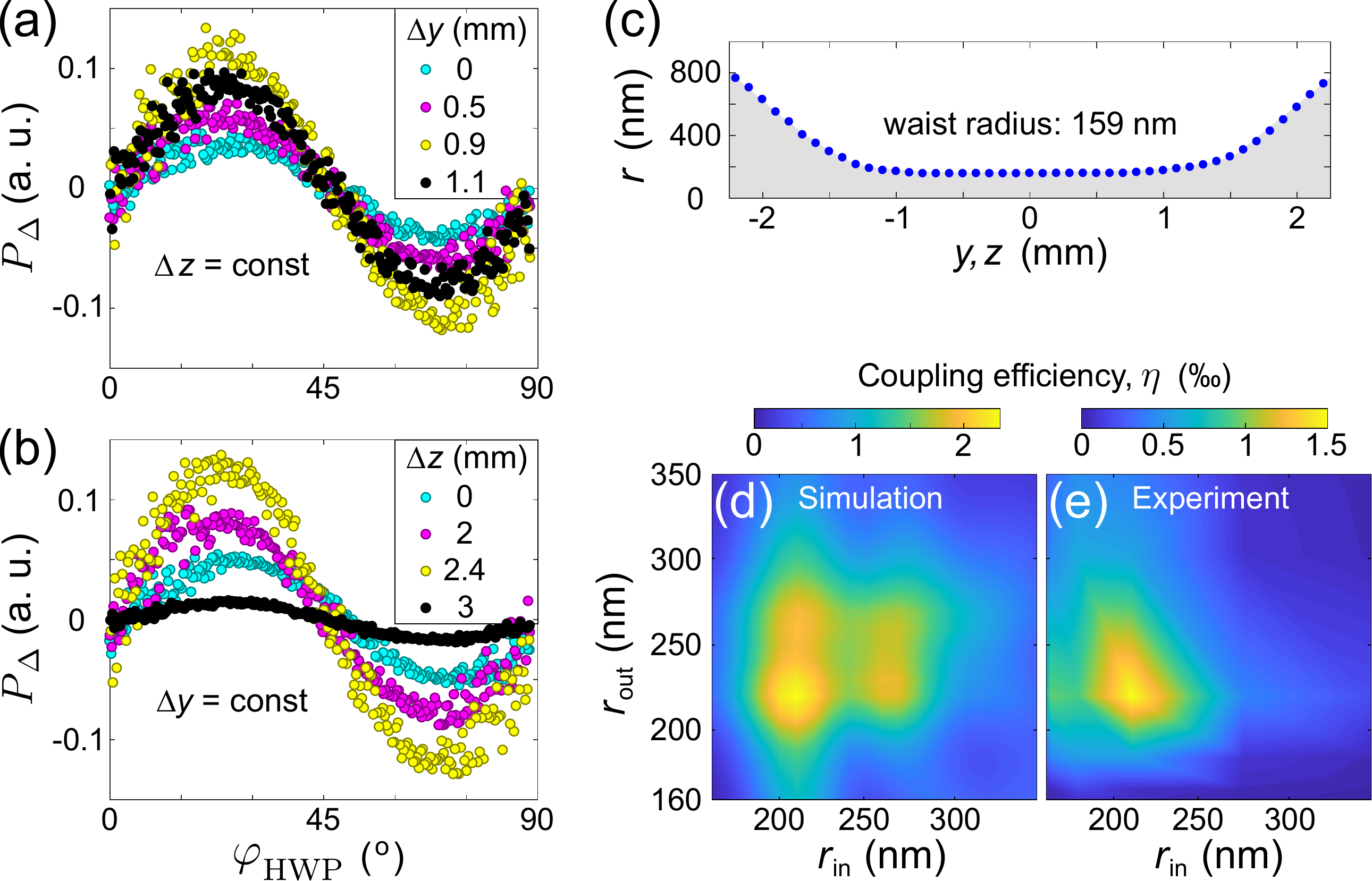}
\caption{The role of crossing-point position and fiber thickness. (a)~Output power difference, $P_{\Delta}$, measured at a fixed point $(0,0,\Delta z)$ on the input fiber with the output fiber being displaced vertically by $\Delta y$. Although the amplitude of $P_{\Delta}$ depends on $\Delta y$, its shape and phase are preserved. Therefore, polarization information gathered by the output fiber does not depend on its radius. (b)~The input fiber is probed at various $\Delta z$ by the output fiber of a fixed radius. The behaviour of $P_{\Delta}$ indicates that the polarization state is maintained throughout the whole coupling region. (c)~A typical nanofiber radius profile measured by a scanning electron microscope. (d),(e)~Simulated and measured efficiency of the directional coupler, $\eta = P_{\Sigma}/P_{\rm T}$, dependent on the radii of the nanofibers.}
\label{fig:size}
\end{figure}

In practice, to identify two non-orthogonal polarization states at the waist of a nanofiber, we cross it with a near-identical nanofiber at right angles, as sketched in  Fig.~\ref{fig:coupler}(a). Near-field probing one (input) ultrathin fiber with another (output) one is not new, see for instance its application for the purpose of profilometry~\cite{madsen_nl_2016,fatemi_o_2017}. In this work, for the first time, we consider symmetry of such a system (hence the importance of the right-angle crossing) and use output signals from both ends of the probe fiber. The efficiency of near-field coupling to guided modes which produce these signals is very low, typically under $0.1\%$, see Fig.~\ref{fig:size}(d),(e). However, it is sufficient for unambiguous identification of non-orthogonal $\bf H$ and ${\bf R}_{\rm M}$ (see below) polarization states at the waist of the input fiber.

The system of two crossed fibers has a symmetry plane, $(x,z)$, shown as the dash-dotted line in Figs.~\ref{fig:coupler}(b),(c). When the input beam is $\bf H$- or $\bf V$-polarized, the overall symmetry of the light-matter system is preserved and the optical power values, $P_+$ and $P_-$, measured at the ends of the output fiber, are equal: $P_+=P_-$. Otherwise, the symmetry is broken and in general $P_+\neq P_-$. Interestingly, this directional coupling of light can be separated into two independent effects associated with orientation ($\psi$) and shape ($\chi$) of the polarization ellipse. Let us now consider the two special cases of linear ($\psi\neq{\rm const}$; $\chi=0$) and circular ($\chi=\pm\pi/4$) polarizations.

For a tilted linear polarization (Fig.~\ref{fig:coupler}(b)), the mirror symmetry is broken by the emission pattern of an electric dipole induced at the crossing point. As a result, two counter-propagating modes of non-equal power are generated in the output fiber. A wave with reflection angle, $\theta$, carries energy proportional to $\sin^2(\theta+\psi)$. Assuming that all angles allowed in this nanofiber equally contribute to the energy transfer, the net power radiated along the $\pm$ direction is proportional to $\int_{\theta_{\rm c}}^{\pi/2}\sin^2(\theta\pm\psi)d\theta$, where $\theta_{\rm c}$ is the critical angle. Therefore, the output power sum,
\begin{equation}
  P_{\Sigma}=(P_++P_-)\propto(\cos2\psi+{\rm const})\,,
  \label{eq:p_sigma}
\end{equation}
which has a maximum (minimum) for ${\bf H}$ (${\bf V}$) polarization. This effect, which we dubbed ``asymmetric dipolar emission'' in order to emphasize its geometrical origin, provides an alternative way to achieve directionality without spin-momentum locking~\cite{picardi_prl_2018}. This possibility has been overlooked in earlier works on directional coupling in similar systems~\cite{petersen_s_2014,sadgrove_sr_2017}.

For circular polarization, the electric field vector, ${\bf E}(t)$, traces a circle about the longitudinal axis, $z$, see Fig.~\ref{fig:coupler}(c). In free space, such a field only contains the longitudinal component of the photon spin, ${\bf J}_z$. The evanescent field may also contain a significant transverse spin component, ${\bf J}_{\rm trans}$, which appears due to interaction between the real, ${\rm Re}({\bf k})$, and imaginary, ${\rm Im}({\bf k})$, parts of the wave vector, $\bf k$, at the interface between two different media~\cite{bliokh_np_2015}. Interestingly, ${\rm Re}({\bf k})$, ${\rm Im}({\bf k})$, and ${\bf J}_{\rm trans}$ must form a right-handed system~\cite{vanMechelen_o_2016}. As a result, the direction of ${\bf J}_{\rm trans}$ defines the direction of the propagating wave~\cite{petersen_s_2014,oconnor_nc_2014}. This phenomenon is known as ``spin locking'' or the ``quantum spin Hall effect'' of light~\cite{bliokh_s_2015}. With regard to our crossed nanofibers, spin locking causes $P_-\neq P_+$, since the vector ${\bf J}_z$ is simultaneously the longitudinal spin for the input nanofiber and the transverse spin for one of the counter-propagating, $x$-polarized modes of the output nanofiber. Simple trigonometric considerations yield that the output power difference produced by the spin locking effect,
\begin{equation}
  P_{\Delta}=(P_+-P_-)\propto \sin2\chi\,,
  \label{eq:p_delta}
\end{equation}
with maximum and minimum values at $\chi=\pi/4$ (${\bf R}$ polarization state) and $\chi=-\pi/4$ (${\bf L}$ state), respectively.

\begin{figure}
\centering
\includegraphics[width=1\linewidth]{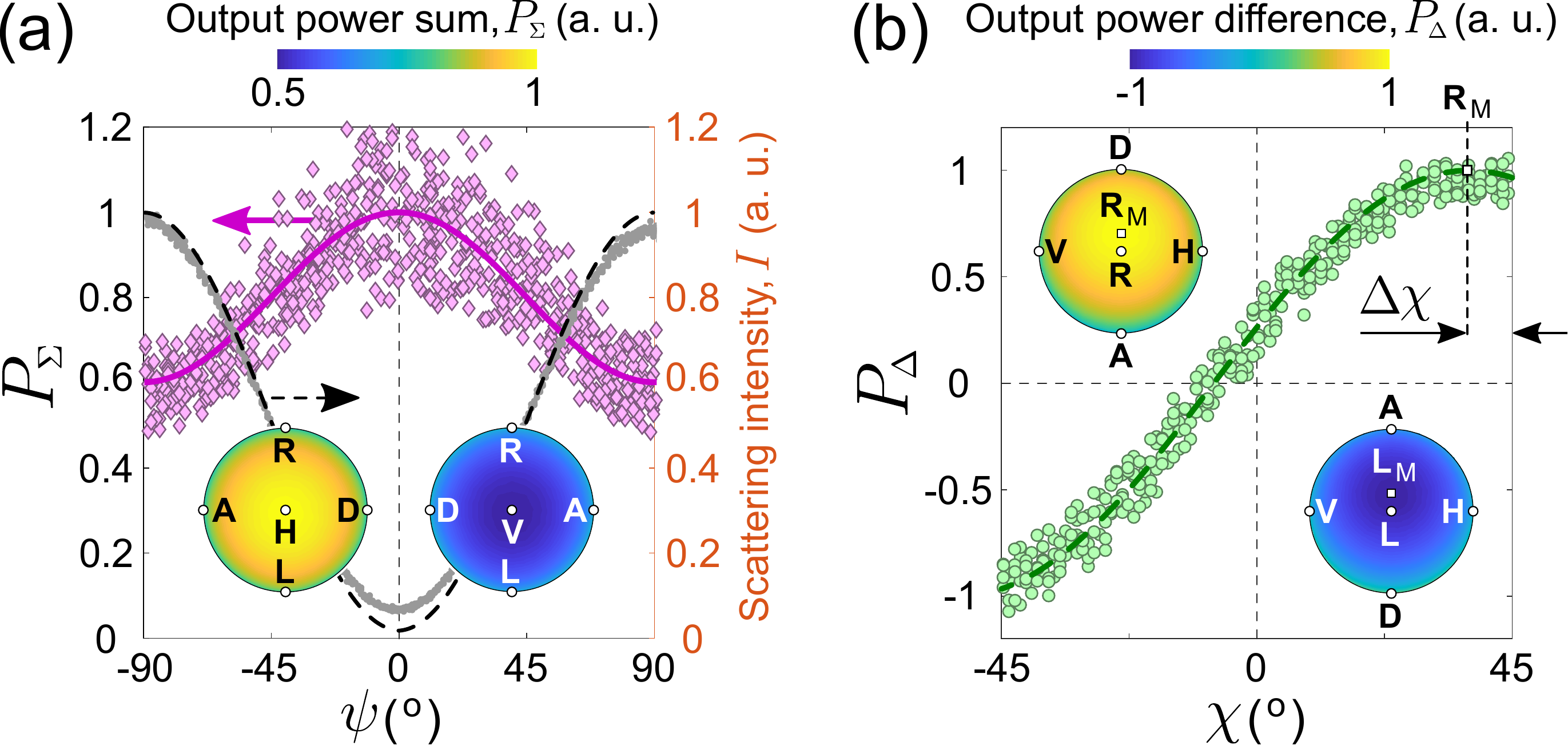}
\caption{Operation of the crossed-nanofiber coupler. (a)~Measured (diamonds) and simulated (solid curve) values of the output power sum, $P_{\Sigma}$, versus orientation of the linear polarization, $\psi$. The numerical simulation over the Poincar\'e sphere (insets) shows that $P_{\Sigma}$ has the global maximum for the horizontal and the global minimum for the vertical polarization states. The intensity of scattering from the fiber crossing (dots---measured, dashed curve---simulated) is exactly opposite in phase with respect to $P_{\Sigma}$. (b)~Measured (circles) and simulated (dashed curve) values of the output power difference, $P_{\Delta}$, versus the polarization ellipticity. The simulated global maximum corresponds to the state ${\bf R}_{\rm M}$, which is shifted from ${\bf R}$ towards ${\bf D}$ due to the interplay between the two mechanisms of mirror symmetry breaking.}
\label{fig:operation}
\end{figure}

For  elliptical polarization ($\psi,\chi\neq0$), both asymmetric dipolar emission and spin locking effects may influence the output power balance. In order to confirm the above analytical predictions for the special cases and generalize the understanding of the directional coupler's operation, finite-element numerical simulations of $P_\pm$ for ${\bf s}_{\rm w}$ covering the whole Poincar\'e sphere were performed. The results for variable $\psi$ and $\chi$ are shown in Figs.~\ref{fig:operation}(a),(b)~\footnote{For the ease of presentation, this figure shows experimental data collected after the polarization compensation.}. The insets in Fig.~\ref{fig:operation}(a) demonstrate that $\bf H$ and $\bf V$ correspond to the global extrema of $P_{\Sigma}$ and  each of these two states can be identified and used in the first step of the polarization compensation. Experimentally, $\psi$ was varied while keeping $S_3=0$ by sending the $\bf H$-polarized input beam through a rotating half-wave plate (HWP). The data (diamonds) agree with the simulation (solid curve). Thus, $P_{\Sigma}$ can be readily used for ${\bf H}\to{\bf H}$ mapping, which can be further verified by monitoring the intensity of scattering from the fiber crossing, $I$, see the experimental (dots) and theoretical (dashed curve) results in Fig.~\ref{fig:operation}(a). We measured $I$ as the total brightness of the camera image captured through a linear polarizer parallel to the $y$ axis~\cite{vetsch_ieee_2012}.

\begin{figure}
\centering
\includegraphics[width=0.9\linewidth]{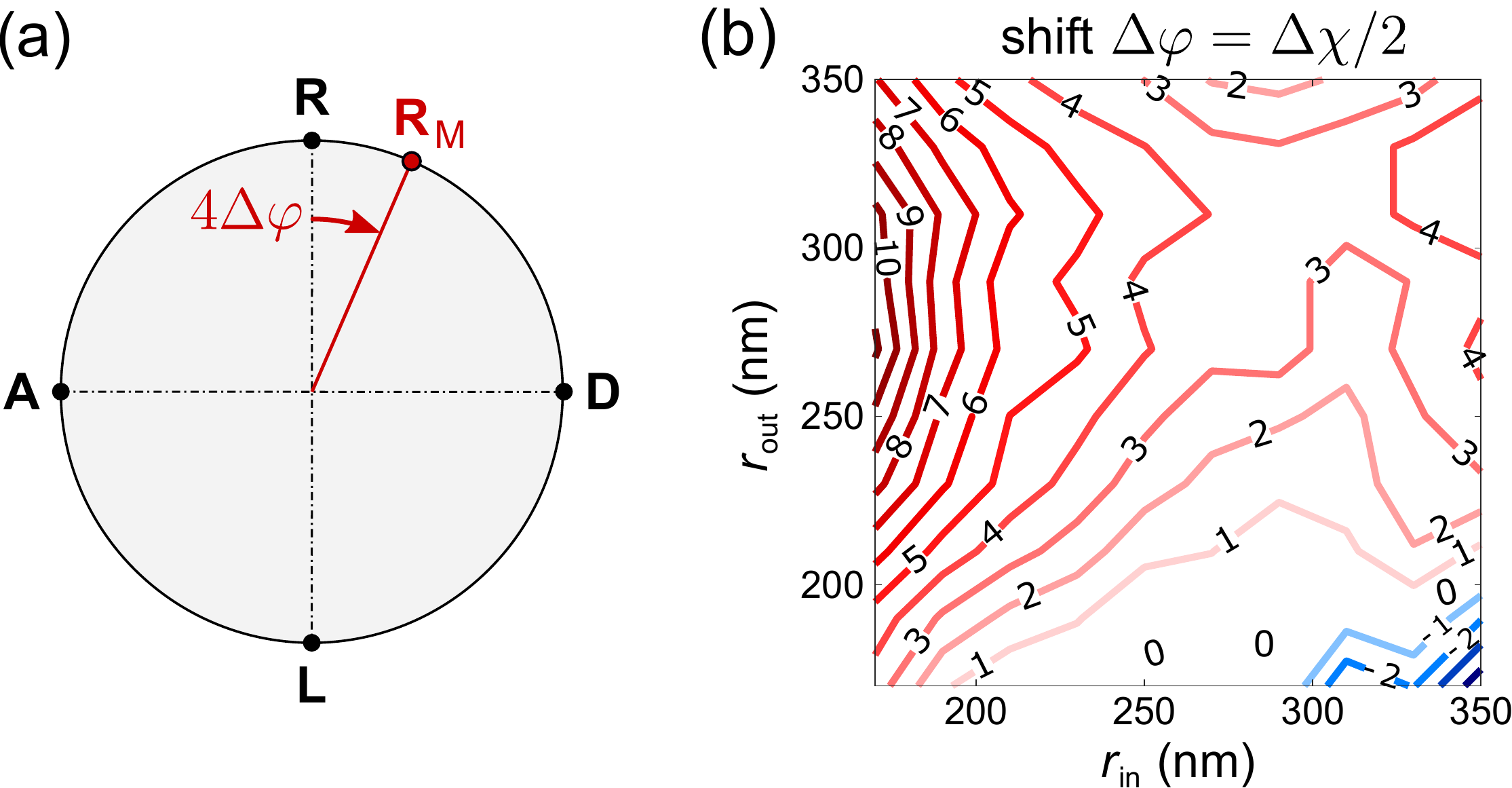}
\caption{Shift of the maximum $P_{\Delta}$ for ${\bf s}_{\rm in}$ tracing the circle in the $S_1=0$ plane of the Poincar\'e sphere. (a)~On the sphere, the shift appears as $2\Delta\chi=4\Delta\varphi$, where $\Delta\varphi$ is the corresponding orientation of the HWP in the polarization generator. (b)~Simulated map of $\Delta\varphi$ versus radii of the input ($r_{\rm in}$) and output ($r_{\rm out}$) nanofibers. Noteworthy, the shift can be negative, in which case ${\bf R}_{\rm M}$ is located between $\bf R$ and~$\bf A$. However, in practice, the coupling efficiency of the crossed fibers in this regime is too weak to be detected, see Figs.~\ref{fig:size}(d),(e)}
\label{fig:shift}
\end{figure}

The simulations of $P_{\Delta}$ (see  insets 
in Fig.~\ref{fig:operation}(b)) reveal that, due to the interplay between the asymmetric dipolar emission and the spin locking effects, the global maximum (minimum) of $P_{\Delta}$ is shifted from the expected $\bf R$ ($\bf L$) to ${\bf R}_{\rm M}$ (${\bf L}_{\rm M}$) by $\Delta\chi\approx8^{\circ}$. This value depends on the radii of the nanofibers and was under $10^{\circ}$ for the region where significant coupling can be achieved (see Fig.~\ref{fig:shift}(b) and Figs.~\ref{fig:size}(d),(e)). Experimentally, $\chi$ was varied while keeping $S_1=0$ by rotation of a HWP in front of a QWP fixed at $45^{\circ}$ to the $x$ axis, as depicted in Fig.~\ref{fig:accuracy}(g). In order to avoid systematic errors associated with $\Delta\chi$, in the second step of the polarization control, we mapped ${\bf R}_{\rm M}\to{\bf R}_{\rm M}$ instead of ${\bf R}\to{\bf R}$.

All data presented in Fig.~\ref{fig:operation} were collected while keeping the fiber crossing point fixed. Now let us check whether the choice of the crossing point makes a difference. For instance, if the polarization at the input fiber waist depends on the longitudinal position, $z$, the curves for $P_{\Sigma,\Delta}$ versus the HWP orientation, $\varphi_{\rm HWP}$, will have variable phase shifts. Alterations to $P_{\Sigma,\Delta}$ will also appear if the directional coupling depends on the vertical position of the output fiber, or, effectively, its radius at the crossing. We tested both possibilities using one period of the ${\bf HRVL}$ trajectory on the Poincar\'e sphere (Figs.~\ref{fig:accuracy}(d),(f)). First, we displaced the output fiber along its axis, thus varying $\Delta y = (y-y_0)$ while keeping the longitudinal position of the crossing point fixed, so that $\Delta z = (z-z_0) = {\rm const}$, where $y_0$,\,$z_0$ are random initial coordinates. As shown in Fig.~\ref{fig:size}(a), the measured $P_{\Delta}$ is independent of $\Delta y$, up to an amplitude scaling factor. The same behavior was observed when the crossing point was displaced along $z$ with $\Delta y = {\rm const}$, see Fig.~\ref{fig:size}(b). These results indicate that ${\bf s}_{\rm w}$ is maintained throughout the coupling region, which we define as the $y\times z$ area where light can still couple to the output fiber. We found that the amplitudes of $P_+$ and $P_-$ are nonzero within $4$~mm$\times4$~mm area covering the waist and thinner parts of the tapers, see the radius profile in Fig.~\ref{fig:size}(c).

According to our numerical simulations presented in Fig.~\ref{fig:size}(d), the coupling efficiency, $\eta = P_{\Sigma}/P_{\rm T}$ (where $P_{\rm T}$ is the optical power transmitted through the input fiber) is maximum for an input nanofiber radius, $r_{\rm in}^{\rm max}\approx210$~nm, and an output nanofiber radius, $r_{\rm out}^{\rm max}\approx220$~nm. We have measured $\eta$ over the whole coupling range with a step of 0.25~mm in $y$ and $z$. The resulting efficiency map shown in Fig.~\ref{fig:size}(e) agrees reasonably with the simulations. 
Notably, the whole explored range for the nanofiber radii, $r$, corresponds to the single-mode regime, since the normalized frequency parameter, $V = (2\pi r/\lambda) \sqrt{1.45^2-1}<2.356$, is below the cut-off value of 2.405~\cite{snyder_book}. In fact, the polarization compensation is valid only for single-mode input nanofibers, with a minimum wavelength-to-diameter ratio of $\pi$NA$/2.405$ (about 1.372 for glass in air).

\section{V. Polarization control}
\begin{figure}[b]
\centering
\includegraphics[width=0.95\linewidth]{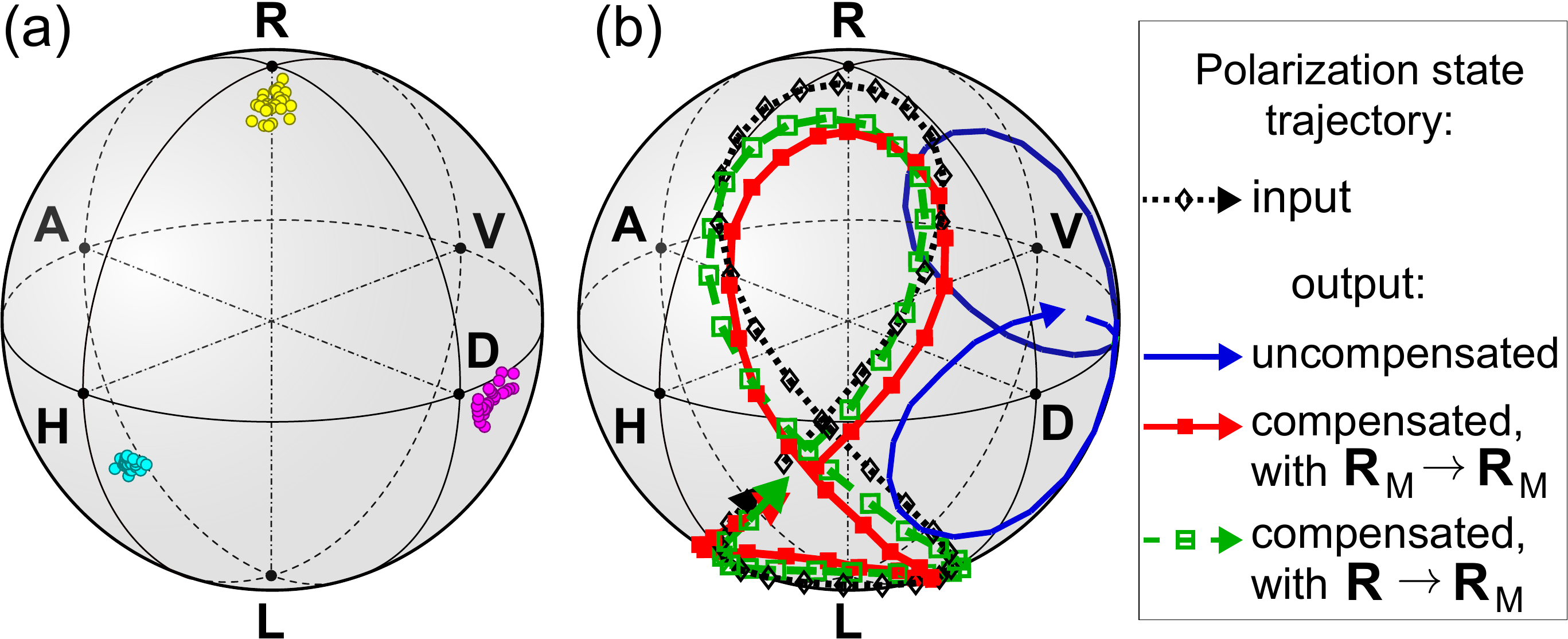}
\caption{Precision of the polarization control. (a)~Statistics for the principal states measured at the end of the input fiber after compensation for a random~${\rm M}_-$. (b)~Recovering an arbitrary input trajectory (black diamonds) for ${\bf s}_{\rm w}$ using compensation with ${\bf R}_{\rm M}\to{\bf R}_{\rm M}$ (solid red squares) or ${\bf R}\to{\bf R}_{\rm M}$ (empty green squares) mapping.}
\label{fig:precision}
\end{figure}

Figure \ref{fig:precision} illustrates the {\it precision} of the polarization control we achieved.  In order to estimate the precision, the input pigtail was spliced to fiber paddles (Fig.~\ref{fig:concept}(c)) that allowed us to produce a random~${\rm M}_-$. For each random setting of the paddles,  the two-step compensation procedure was performed by ${\bf H}\to{\bf H}$ and ${\bf R}_{\rm M}\to{\bf R}_{\rm M}$ mapping. Then, ${\bf s}_{\rm out}$ was measured for the three principal states: ${\bf H}$, ${\bf D}$, and ${\bf R}$. The resulting statistics over 26 sets (see Fig.~\ref{fig:precision}(a)) gives the following  deviations from the mean: $1.16\pm1.43^{\circ}$, $6.03\pm3.82^{\circ}$, and $4.29\pm2.37^{\circ}$ for {\bf H}, {\bf D}, and {\bf R}, respectively. The fidelities (defined as the cosine of the mean angular distances) for these states are 0.9998, 0.9945, and~0.9972, respectively. The offsets from the target states (about $10^{\circ}$ for this fiber) are due to the unknown, constant matrix~${\rm M}_+$.

\begin{figure*}
\centering
\includegraphics[width=1\linewidth]{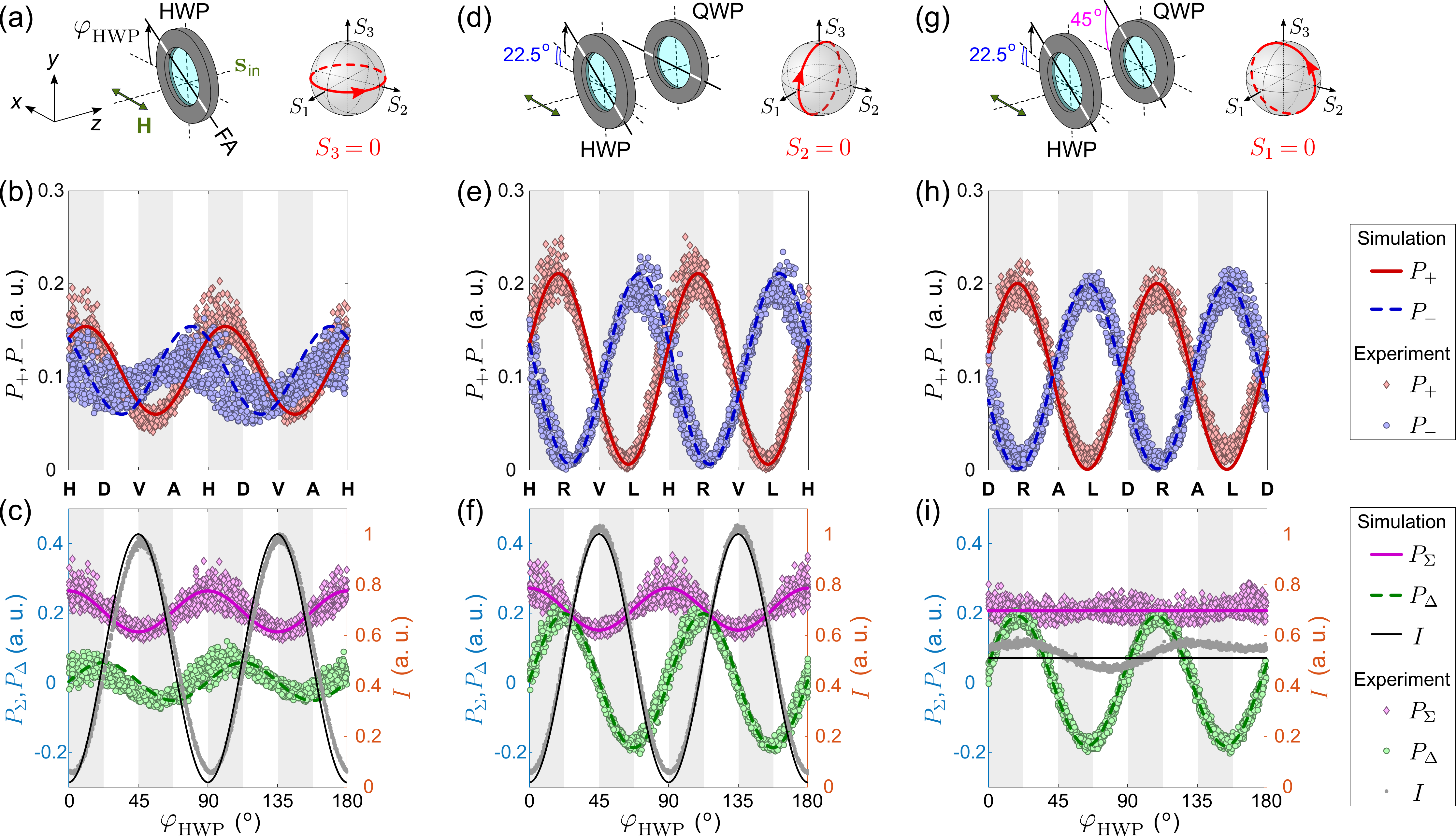}
\caption{Testing the crossed-nanofiber directional coupler after the polarization compensation. (a),(d),(g)~The input polarization state, ${\bf s}_{\rm in}$, traces a circle in one of the principal planes of the Poincar\'e sphere, driven by the polarization generator represented by a rotating HWP and a static QWP. (b)~Simulated (solid and dashed lines) and measured (diamonds and circles) optical powers at the `$+$' and `$-$' ends of the output fiber for ${\bf s}_{\rm in}$ lying in the $S_3=0$ plane. (c)~Simulated (lines) and measured (diamonds and circles) sum and difference of the output powers, along with the scattering intensity (dots). (e),(f)~Same as (b),(c), but for the $S_2=0$ plane. The simulated $P_{\Delta}$ is maximum for the~${\bf R}$ state, and minimum for the~${\bf L}$ state. (h),(i)~Same for the $S_1 = 0$ plane. The simulated maximum of $P_{\Delta}$ is slightly shifted from ${\bf R}$ towards ${\bf D}$ (by $\Delta\varphi$, see Fig.~\ref{fig:shift}) due to the interplay between the asymmetric dipolar emission and the spin locking effects. Here the experimental $P_{-,+,\Sigma,\Delta}$ are in units of voltage from photodetectors; simulated $P_{-,+}$ are normalized by the experimental maxima; scattering intensity $I$ is normalized by its maximum. The grey bands are added as a guide to the eye.}
\label{fig:accuracy}
\end{figure*}

Besides the principal states, the method was tested for a figure-of-eight trajectory (black diamonds and dotted line in Fig.~\ref{fig:nanofiber}(b) and Fig.~\ref{fig:precision}(b)). The randomly arranged paddles could move the uncompensated trajectory to any part of the sphere (see the blue solid curve in Fig.~\ref{fig:precision}(b) as an example). Compensation with ${\bf H}\to{\bf H}$ and ${\bf R}_{\rm M}\to{\bf R}_{\rm M}$ brings the trajectory (red filled squares) close to the initial one. This result indicates that the second (straight) part of the tapered fiber (corresponding to ${\rm M}_+$) contributed only a minor transformation to~${\bf s}_{\rm w}$. The case where the correction to $\Delta\chi$ is ignored was also checked. In this case, the initial state is ${\bf R}$, and ${\bf R}\to{\bf R}_{\rm M}$ mapping is achieved by locating the maximum of $P_{\Delta}$.
This mapping causes systematic errors for $\varphi_2$, see the compensated trajectories in Fig.~\ref{fig:precision}(b).

In order to demonstrate the {\it accuracy} of the polarization control, we have performed a detailed study summarized in Fig.~\ref{fig:accuracy}. In this study, the input polarization state, ${\bf s}_{\rm in}$, was driven along a circular trajectory in one of the principal planes of the Poincar\'e sphere:
$S_3=0$, $S_2=0$, or $S_1 = 0$. This was achieved by the free-space polarization generator depicted in the top panels (that is, (a), (d), and (g)) for each trajectory. The experimental data for the directional coupler's outputs and the scattering intensity were collected after the polarization compensation procedure. We note that spin-locking and asymmetric dipolar emission effects produce comparable modulations of the directionality characterized by $P_{\Delta}$, see Figs.~\ref{fig:accuracy}(c),(f).
The higher noise in the `$-$' channel (especially noticeable in Fig.~\ref{fig:accuracy}(b)) was repeatable over numerous experimental attempts and independent of the detectors. The noise levels in the two channels were closer to each other for thicker nanofibers (with radii over 250~nm). Therefore, we attribute this effect to higher sensitivity of thinner nanofibers to bends due to vibrations or sagging, and, perhaps, to unknown deviations of the generated trajectory from the target one.

Our results contradict the general belief that short lengths of straight tapered fibers do not change the polarization by much and, even if they do, the shifts of the input state, ${\bf s}_{\rm in}$, by the down- and up-tapers are equal, so that the state at the waist, ${\bf s}_{\rm w}$, lies in the center between ${\bf s}_{\rm in}$ and ${\bf s}_{\rm out}$. According to our findings, presented in Fig.~\ref{fig:nanofiber}(b) and Fig.~\ref{fig:precision}(b), the transformation matrices ${\rm M}_-$ and ${\rm M}_+$ lead to oppositely-directed, non-equal shifts of significant magnitudes.
Therefore, it is clear that  experiments with high levels of precision (like those involving quantum emitters) do require near-field polarization control, which can now be achieved using the reported method.

\section{VI. Conclusion}
We demonstrated the first method for the complete control of polarization of light in a single-mode optical nanofiber waveguide, using directional coupling between two such nanofibers.
Although based on complex physical phenomena (the spin-momentum locking and the newly discovered asymmetric dipolar emission), our method is surprisingly simple and highly reliable.
We believe that it will have significant impact on the vast range of experimental systems based on optical nanofibers and evanescently coupled elements, in general. The demonstrated directional coupler itself is a promising platform for developing integrated photonic circuits.

\section*{Acknowledgments}
The authors would like to thank S.~P.~Mekhail for help with automation of the experiment, K.~Karlsson for assistance with nanofiber fabrication, and F.~Le~Kien for helpful discussions. This study was partially supported by the Okinawa Institute of Science and Technology Graduate University. G.~T.~was supported by the Japan Society for the Promotion of Science (JSPS) as an International Research Fellow (Standard, ID~No.P18367). 

\section{A\lowercase{ppendix: polarization compensation requires non-orthogonal states}}

Our polarization compensation is based on sequential mapping of two input polarization states, $\bf A$ and $\bf B$, onto themselves at the waist of an optical nanofiber. As a result, 
\begin{equation}
\begin{cases}
    u{\bf A}={\mathrm e}^{i\delta_A}{\bf A}\\
    u{\bf B}={\mathrm e}^{i\delta_B}{\bf B}
\end{cases}\,,
\end{equation}
where the unknown phase factors, $\delta_A$ and $\delta_B$, do not change the state, and the Jones matrix, $u=u_{\rm fiber}u_{\rm PC}$, describes the polarization transformation in the compensator ($u_{\rm PC}$) and the tapered fiber until the waist ($u_{\rm fiber}$). Since $u$ is unitary, the inner product of these states
\begin{equation}
\bra{{\bf A}}u^{\dagger}u\ket{{\bf B}}=\bra{{\bf A}}\ket{{\bf B}}={\mathrm e}^{i(\delta_A-\delta_B)}\bra{{\bf A}}\ket{{\bf B}}\,.
\label{eq:AB}
\end{equation}
If these two states are not orthogonal, that is
\begin{equation}
\bra{{\bf A}}{\bf B}\rangle\neq 0\,,
\end{equation}
 then Eq.~\ref{eq:AB} requires $\delta_A=\delta_B=\delta$. Under this condition, an arbitrary input state, $\bf C$, can be defined in the basis ($\bf A$,~$\bf B$):
\begin{equation}
\begin{cases}
    {\bf C} = a{\bf A}+b{\bf B}\\
    u{\bf C}={\mathrm e}^{i\delta}{\bf C}
\end{cases}\,,
\end{equation}
where $a$ and $b$ are complex coefficients. Such a state $\bf C$ will be maintained in the polarization-compensated nanofiber.

\bibliography{biblio_pol_control}

\begin{thebibliography}{37}%
\makeatletter
\providecommand \@ifxundefined [1]{%
 \@ifx{#1\undefined}
}%
\providecommand \@ifnum [1]{%
 \ifnum #1\expandafter \@firstoftwo
 \else \expandafter \@secondoftwo
 \fi
}%
\providecommand \@ifx [1]{%
 \ifx #1\expandafter \@firstoftwo
 \else \expandafter \@secondoftwo
 \fi
}%
\providecommand \natexlab [1]{#1}%
\providecommand \enquote  [1]{``#1''}%
\providecommand \bibnamefont  [1]{#1}%
\providecommand \bibfnamefont [1]{#1}%
\providecommand \citenamefont [1]{#1}%
\providecommand \href@noop [0]{\@secondoftwo}%
\providecommand \href [0]{\begingroup \@sanitize@url \@href}%
\providecommand \@href[1]{\@@startlink{#1}\@@href}%
\providecommand \@@href[1]{\endgroup#1\@@endlink}%
\providecommand \@sanitize@url [0]{\catcode `\\12\catcode `\$12\catcode
  `\&12\catcode `\#12\catcode `\^12\catcode `\_12\catcode `\%12\relax}%
\providecommand \@@startlink[1]{}%
\providecommand \@@endlink[0]{}%
\providecommand \url  [0]{\begingroup\@sanitize@url \@url }%
\providecommand \@url [1]{\endgroup\@href {#1}{\urlprefix }}%
\providecommand \urlprefix  [0]{URL }%
\providecommand \Eprint [0]{\href }%
\providecommand \doibase [0]{http://dx.doi.org/}%
\providecommand \selectlanguage [0]{\@gobble}%
\providecommand \bibinfo  [0]{\@secondoftwo}%
\providecommand \bibfield  [0]{\@secondoftwo}%
\providecommand \translation [1]{[#1]}%
\providecommand \BibitemOpen [0]{}%
\providecommand \bibitemStop [0]{}%
\providecommand \bibitemNoStop [0]{.\EOS\space}%
\providecommand \EOS [0]{\spacefactor3000\relax}%
\providecommand \BibitemShut  [1]{\csname bibitem#1\endcsname}%
\let\auto@bib@innerbib\@empty
\bibitem [{\citenamefont {Brambilla}\ \emph {et~al.}(2007)\citenamefont
  {Brambilla}, \citenamefont {{Senthil Murugan}}, \citenamefont {Wilkinson},\
  and\ \citenamefont {Richardson}}]{brambilla_ol_2007}%
  \BibitemOpen
  \bibfield  {author} {\bibinfo {author} {\bibfnamefont {G.}~\bibnamefont
  {Brambilla}}, \bibinfo {author} {\bibfnamefont {G.}~\bibnamefont {{Senthil
  Murugan}}}, \bibinfo {author} {\bibfnamefont {J.~S.}\ \bibnamefont
  {Wilkinson}}, \ and\ \bibinfo {author} {\bibfnamefont {D.~J.}\ \bibnamefont
  {Richardson}},\ }\bibfield  {title} {\enquote {\bibinfo {title} {Optical
  manipulation of microspheres along a subwavelength optical wire},}\
  }\href@noop {} {\bibfield  {journal} {\bibinfo  {journal} {Opt. Lett.}\
  }\textbf {\bibinfo {volume} {32}},\ \bibinfo {pages} {3041--3043} (\bibinfo
  {year} {2007})}\BibitemShut {NoStop}%
\bibitem [{\citenamefont {Maimaiti}\ \emph {et~al.}(2015)\citenamefont
  {Maimaiti}, \citenamefont {Truong}, \citenamefont {Sergides}, \citenamefont
  {Gusachenko},\ and\ \citenamefont {{Nic Chormaic}}}]{maimaiti_sr_2015}%
  \BibitemOpen
  \bibfield  {author} {\bibinfo {author} {\bibfnamefont {A.}~\bibnamefont
  {Maimaiti}}, \bibinfo {author} {\bibfnamefont {V.~G.}\ \bibnamefont
  {Truong}}, \bibinfo {author} {\bibfnamefont {M.}~\bibnamefont {Sergides}},
  \bibinfo {author} {\bibfnamefont {I.}~\bibnamefont {Gusachenko}}, \ and\
  \bibinfo {author} {\bibfnamefont {S.}~\bibnamefont {{Nic Chormaic}}},\
  }\bibfield  {title} {\enquote {\bibinfo {title} {Higher order microfibre
  modes for dielectric particle trapping and propulsion},}\ }\href@noop {}
  {\bibfield  {journal} {\bibinfo  {journal} {Sci. Rep.}\ }\textbf {\bibinfo
  {volume} {5}},\ \bibinfo {pages} {9077} (\bibinfo {year} {2015})}\BibitemShut
  {NoStop}%
\bibitem [{\citenamefont {Ren}\ \emph {et~al.}(2016)\citenamefont {Ren},
  \citenamefont {Zhang}, \citenamefont {Ti},\ and\ \citenamefont
  {Liu}}]{ren2016tapered}%
  \BibitemOpen
  \bibfield  {author} {\bibinfo {author} {\bibfnamefont {Y.}~\bibnamefont
  {Ren}}, \bibinfo {author} {\bibfnamefont {R.}~\bibnamefont {Zhang}}, \bibinfo
  {author} {\bibfnamefont {C.}~\bibnamefont {Ti}}, \ and\ \bibinfo {author}
  {\bibfnamefont {Y.}~\bibnamefont {Liu}},\ }\bibfield  {title} {\enquote
  {\bibinfo {title} {Tapered optical fiber loops and helices for integrated
  photonic device characterization and microfluidic roller coasters},}\
  }\href@noop {} {\bibfield  {journal} {\bibinfo  {journal} {Optica}\ }\textbf
  {\bibinfo {volume} {3}},\ \bibinfo {pages} {1205--1208} (\bibinfo {year}
  {2016})}\BibitemShut {NoStop}%
\bibitem [{\citenamefont {Yoshie}\ \emph {et~al.}(2011)\citenamefont {Yoshie},
  \citenamefont {Tang},\ and\ \citenamefont {Su}}]{yoshie_sensors_2011}%
  \BibitemOpen
  \bibfield  {author} {\bibinfo {author} {\bibfnamefont {T.}~\bibnamefont
  {Yoshie}}, \bibinfo {author} {\bibfnamefont {L.}~\bibnamefont {Tang}}, \ and\
  \bibinfo {author} {\bibfnamefont {S.-Y.}\ \bibnamefont {Su}},\ }\bibfield
  {title} {\enquote {\bibinfo {title} {Optical microcavity: sensing down to
  single molecules and atoms},}\ }\href@noop {} {\bibfield  {journal} {\bibinfo
   {journal} {Sensors}\ }\textbf {\bibinfo {volume} {11}},\ \bibinfo {pages}
  {1972--1991} (\bibinfo {year} {2011})}\BibitemShut {NoStop}%
\bibitem [{\citenamefont {Yu}\ \emph {et~al.}(2014)\citenamefont {Yu},
  \citenamefont {Li}, \citenamefont {Wang}, \citenamefont {Tong}, \citenamefont
  {Jiang}, \citenamefont {Li}, \citenamefont {Gong},\ and\ \citenamefont
  {Xiao}}]{yu_am_2014}%
  \BibitemOpen
  \bibfield  {author} {\bibinfo {author} {\bibfnamefont {X.-C.}\ \bibnamefont
  {Yu}}, \bibinfo {author} {\bibfnamefont {B.-B.}\ \bibnamefont {Li}}, \bibinfo
  {author} {\bibfnamefont {P.}~\bibnamefont {Wang}}, \bibinfo {author}
  {\bibfnamefont {L.}~\bibnamefont {Tong}}, \bibinfo {author} {\bibfnamefont
  {X.-F.}\ \bibnamefont {Jiang}}, \bibinfo {author} {\bibfnamefont
  {Y.}~\bibnamefont {Li}}, \bibinfo {author} {\bibfnamefont {Q.}~\bibnamefont
  {Gong}}, \ and\ \bibinfo {author} {\bibfnamefont {Y.-F.}\ \bibnamefont
  {Xiao}},\ }\bibfield  {title} {\enquote {\bibinfo {title} {Single
  nanoparticle detection and sizing using a nanofiber pair in an aqueous
  environment},}\ }\href@noop {} {\bibfield  {journal} {\bibinfo  {journal}
  {Adv. Mater.}\ }\textbf {\bibinfo {volume} {26}},\ \bibinfo {pages}
  {7462--7467} (\bibinfo {year} {2014})}\BibitemShut {NoStop}%
\bibitem [{\citenamefont {Beugnot}\ \emph {et~al.}(2014)\citenamefont
  {Beugnot}, \citenamefont {Lebrun}, \citenamefont {Pauliat}, \citenamefont
  {Maillotte}, \citenamefont {Laude},\ and\ \citenamefont
  {Sylvestre}}]{beugnot2014brillouin}%
  \BibitemOpen
  \bibfield  {author} {\bibinfo {author} {\bibfnamefont {J.-C.}\ \bibnamefont
  {Beugnot}}, \bibinfo {author} {\bibfnamefont {S.}~\bibnamefont {Lebrun}},
  \bibinfo {author} {\bibfnamefont {G.}~\bibnamefont {Pauliat}}, \bibinfo
  {author} {\bibfnamefont {H.}~\bibnamefont {Maillotte}}, \bibinfo {author}
  {\bibfnamefont {V.}~\bibnamefont {Laude}}, \ and\ \bibinfo {author}
  {\bibfnamefont {T.}~\bibnamefont {Sylvestre}},\ }\bibfield  {title} {\enquote
  {\bibinfo {title} {Brillouin light scattering from surface acoustic waves in
  a subwavelength-diameter optical fibre},}\ }\href@noop {} {\bibfield
  {journal} {\bibinfo  {journal} {Nat.~Commun.}\ }\textbf {\bibinfo {volume}
  {5}},\ \bibinfo {pages} {5242} (\bibinfo {year} {2014})}\BibitemShut
  {NoStop}%
\bibitem [{\citenamefont {{Zoubi and K.~Hammerer}}(2017)}]{zoubi_prl_2017}%
  \BibitemOpen
  \bibfield  {author} {\bibinfo {author} {\bibfnamefont {H.}~\bibnamefont
  {{Zoubi and K.~Hammerer}}},\ }\bibfield  {title} {\enquote {\bibinfo {title}
  {Quantum nonlinear optics in optomechanical nanoscale waveguides},}\ }\href
  {\doibase 10.1103/PhysRevLett.119.123602} {\bibfield  {journal} {\bibinfo
  {journal} {Phys. Rev. Lett.}\ }\textbf {\bibinfo {volume} {119}},\ \bibinfo
  {pages} {123602} (\bibinfo {year} {2017})}\BibitemShut {NoStop}%
\bibitem [{\citenamefont {Cai}\ \emph {et~al.}(2000)\citenamefont {Cai},
  \citenamefont {Painter},\ and\ \citenamefont {Vahala}}]{cai_prl_2000}%
  \BibitemOpen
  \bibfield  {author} {\bibinfo {author} {\bibfnamefont {M.}~\bibnamefont
  {Cai}}, \bibinfo {author} {\bibfnamefont {O.}~\bibnamefont {Painter}}, \ and\
  \bibinfo {author} {\bibfnamefont {K.~J.}\ \bibnamefont {Vahala}},\ }\bibfield
   {title} {\enquote {\bibinfo {title} {Observation of critical coupling in a
  fiber taper to a silica-microsphere whispering-gallery mode system},}\
  }\href@noop {} {\bibfield  {journal} {\bibinfo  {journal} {Phys. Rev. Lett.}\
  }\textbf {\bibinfo {volume} {85}},\ \bibinfo {pages} {74} (\bibinfo {year}
  {2000})}\BibitemShut {NoStop}%
\bibitem [{\citenamefont {{Le Kien}}\ \emph {et~al.}(2004)\citenamefont {{Le
  Kien}}, \citenamefont {Balykin},\ and\ \citenamefont {Hakuta}}]{le2004atom}%
  \BibitemOpen
  \bibfield  {author} {\bibinfo {author} {\bibfnamefont {F.}~\bibnamefont {{Le
  Kien}}}, \bibinfo {author} {\bibfnamefont {V.~I.}\ \bibnamefont {Balykin}}, \
  and\ \bibinfo {author} {\bibfnamefont {K.}~\bibnamefont {Hakuta}},\
  }\bibfield  {title} {\enquote {\bibinfo {title} {Atom trap and waveguide
  using a two-color evanescent light field around a subwavelength-diameter
  optical fiber},}\ }\href@noop {} {\bibfield  {journal} {\bibinfo  {journal}
  {Phys. Rev. A}\ }\textbf {\bibinfo {volume} {70}},\ \bibinfo {pages} {063403}
  (\bibinfo {year} {2004})}\BibitemShut {NoStop}%
\bibitem [{\citenamefont {Hendrickson}\ \emph {et~al.}(2010)\citenamefont
  {Hendrickson}, \citenamefont {Lai}, \citenamefont {Pittman},\ and\
  \citenamefont {Franson}}]{hendrickson_prl_2010}%
  \BibitemOpen
  \bibfield  {author} {\bibinfo {author} {\bibfnamefont {S.~M.}\ \bibnamefont
  {Hendrickson}}, \bibinfo {author} {\bibfnamefont {M.~M.}\ \bibnamefont
  {Lai}}, \bibinfo {author} {\bibfnamefont {T.~B.}\ \bibnamefont {Pittman}}, \
  and\ \bibinfo {author} {\bibfnamefont {J.~D.}\ \bibnamefont {Franson}},\
  }\bibfield  {title} {\enquote {\bibinfo {title} {Observation of two-photon
  absorption at low power levels using tapered optical fibers in rubidium
  vapor},}\ }\href@noop {} {\bibfield  {journal} {\bibinfo  {journal} {Phys.
  Rev. Lett.}\ }\textbf {\bibinfo {volume} {105}},\ \bibinfo {pages} {173602}
  (\bibinfo {year} {2010})}\BibitemShut {NoStop}%
\bibitem [{\citenamefont {Vetsch}\ \emph {et~al.}(2012)\citenamefont {Vetsch},
  \citenamefont {Dawkins}, \citenamefont {Mitsch}, \citenamefont {Reitz},
  \citenamefont {Schneeweiss},\ and\ \citenamefont
  {Rauschenbeutel}}]{vetsch_ieee_2012}%
  \BibitemOpen
  \bibfield  {author} {\bibinfo {author} {\bibfnamefont {E.}~\bibnamefont
  {Vetsch}}, \bibinfo {author} {\bibfnamefont {S.~T.}\ \bibnamefont {Dawkins}},
  \bibinfo {author} {\bibfnamefont {R.}~\bibnamefont {Mitsch}}, \bibinfo
  {author} {\bibfnamefont {D.}~\bibnamefont {Reitz}}, \bibinfo {author}
  {\bibfnamefont {P.}~\bibnamefont {Schneeweiss}}, \ and\ \bibinfo {author}
  {\bibfnamefont {A.}~\bibnamefont {Rauschenbeutel}},\ }\bibfield  {title}
  {\enquote {\bibinfo {title} {Nanofiber-based optical trapping of cold neutral
  atoms},}\ }\href@noop {} {\bibfield  {journal} {\bibinfo  {journal} {IEEE J.
  Sel. Top. Quantum Electron.}\ }\textbf {\bibinfo {volume} {18}},\ \bibinfo
  {pages} {1763--1771} (\bibinfo {year} {2012})}\BibitemShut {NoStop}%
\bibitem [{\citenamefont {Kumar}\ \emph {et~al.}(2015)\citenamefont {Kumar},
  \citenamefont {Gokhroo},\ and\ \citenamefont {{Nic
  Chormaic}}}]{kumar_njp_2015}%
  \BibitemOpen
  \bibfield  {author} {\bibinfo {author} {\bibfnamefont {R.}~\bibnamefont
  {Kumar}}, \bibinfo {author} {\bibfnamefont {V.}~\bibnamefont {Gokhroo}}, \
  and\ \bibinfo {author} {\bibfnamefont {S.}~\bibnamefont {{Nic Chormaic}}},\
  }\bibfield  {title} {\enquote {\bibinfo {title} {Multi-level cascaded
  electromagnetically induced transparency in cold atoms using an optical
  nanofibre interface},}\ }\href@noop {} {\bibfield  {journal} {\bibinfo
  {journal} {New J. Phys.}\ }\textbf {\bibinfo {volume} {17}},\ \bibinfo
  {pages} {123012} (\bibinfo {year} {2015})}\BibitemShut {NoStop}%
\bibitem [{\citenamefont {Kornovan}\ \emph {et~al.}(2017)\citenamefont
  {Kornovan}, \citenamefont {Petrov},\ and\ \citenamefont
  {Iorsh}}]{kornovan_prb_2018}%
  \BibitemOpen
  \bibfield  {author} {\bibinfo {author} {\bibfnamefont {D.~F.}\ \bibnamefont
  {Kornovan}}, \bibinfo {author} {\bibfnamefont {M.~I.}\ \bibnamefont
  {Petrov}}, \ and\ \bibinfo {author} {\bibfnamefont {I.~V.}\ \bibnamefont
  {Iorsh}},\ }\bibfield  {title} {\enquote {\bibinfo {title} {Transport and
  collective radiance in a basic quantum chiral optical model},}\ }\href@noop
  {} {\bibfield  {journal} {\bibinfo  {journal} {Phys. Rev. B}\ }\textbf
  {\bibinfo {volume} {96}},\ \bibinfo {pages} {115162} (\bibinfo {year}
  {2017})}\BibitemShut {NoStop}%
\bibitem [{\citenamefont {Guo}\ \emph {et~al.}(2009)\citenamefont {Guo},
  \citenamefont {Qiu}, \citenamefont {Bao}, \citenamefont {Wiley},
  \citenamefont {Yang}, \citenamefont {Zhang}, \citenamefont {Ma},
  \citenamefont {Yu},\ and\ \citenamefont {Tong}}]{tong_nl_2009}%
  \BibitemOpen
  \bibfield  {author} {\bibinfo {author} {\bibfnamefont {X.}~\bibnamefont
  {Guo}}, \bibinfo {author} {\bibfnamefont {M.}~\bibnamefont {Qiu}}, \bibinfo
  {author} {\bibfnamefont {J.}~\bibnamefont {Bao}}, \bibinfo {author}
  {\bibfnamefont {B.~J.}\ \bibnamefont {Wiley}}, \bibinfo {author}
  {\bibfnamefont {Q.}~\bibnamefont {Yang}}, \bibinfo {author} {\bibfnamefont
  {X.}~\bibnamefont {Zhang}}, \bibinfo {author} {\bibfnamefont
  {Y.}~\bibnamefont {Ma}}, \bibinfo {author} {\bibfnamefont {H.}~\bibnamefont
  {Yu}}, \ and\ \bibinfo {author} {\bibfnamefont {L.}~\bibnamefont {Tong}},\
  }\bibfield  {title} {\enquote {\bibinfo {title} {Direct coupling of plasmonic
  and photonic nanowires for hybrid nanophotonic components and circuits},}\
  }\href@noop {} {\bibfield  {journal} {\bibinfo  {journal} {Nano Lett.}\
  }\textbf {\bibinfo {volume} {9}},\ \bibinfo {pages} {4515--4519} (\bibinfo
  {year} {2009})}\BibitemShut {NoStop}%
\bibitem [{\citenamefont {Tong}\ \emph {et~al.}(2012)\citenamefont {Tong},
  \citenamefont {Zi}, \citenamefont {Guo},\ and\ \citenamefont
  {Lou}}]{tong_oc_2012}%
  \BibitemOpen
  \bibfield  {author} {\bibinfo {author} {\bibfnamefont {L.}~\bibnamefont
  {Tong}}, \bibinfo {author} {\bibfnamefont {F.}~\bibnamefont {Zi}}, \bibinfo
  {author} {\bibfnamefont {X.}~\bibnamefont {Guo}}, \ and\ \bibinfo {author}
  {\bibfnamefont {J.}~\bibnamefont {Lou}},\ }\bibfield  {title} {\enquote
  {\bibinfo {title} {Optical microfibers and nanofibers: a tutorial},}\
  }\href@noop {} {\bibfield  {journal} {\bibinfo  {journal} {Opt. Commun.}\
  }\textbf {\bibinfo {volume} {285}},\ \bibinfo {pages} {4641--47} (\bibinfo
  {year} {2012})}\BibitemShut {NoStop}%
\bibitem [{\citenamefont {Solano}\ \emph {et~al.}(2017)\citenamefont {Solano},
  \citenamefont {Grover}, \citenamefont {Hoffman}, \citenamefont {Ravets},
  \citenamefont {Fatemi}, \citenamefont {Orozco},\ and\ \citenamefont
  {Rolston}}]{solano_chapter_2017}%
  \BibitemOpen
  \bibfield  {author} {\bibinfo {author} {\bibfnamefont {P.}~\bibnamefont
  {Solano}}, \bibinfo {author} {\bibfnamefont {J.~A.}\ \bibnamefont {Grover}},
  \bibinfo {author} {\bibfnamefont {J.~E.}\ \bibnamefont {Hoffman}}, \bibinfo
  {author} {\bibfnamefont {S.}~\bibnamefont {Ravets}}, \bibinfo {author}
  {\bibfnamefont {F.~K.}\ \bibnamefont {Fatemi}}, \bibinfo {author}
  {\bibfnamefont {L.~A.}\ \bibnamefont {Orozco}}, \ and\ \bibinfo {author}
  {\bibfnamefont {S.~L.}\ \bibnamefont {Rolston}},\ }\href@noop {} {\emph
  {\bibinfo {title} {Advances in atomic, molecular, and optical physics}}}\
  (\bibinfo  {publisher} {Elsevier},\ \bibinfo {year} {2017})\ pp.\ \bibinfo
  {pages} {439--505}\BibitemShut {NoStop}%
\bibitem [{\citenamefont {{Le~Kien}}\ \emph {et~al.}(2004)\citenamefont
  {{Le~Kien}}, \citenamefont {Liang}, \citenamefont {Hakuta},\ and\
  \citenamefont {Balykin}}]{le_kien_oc_2004}%
  \BibitemOpen
  \bibfield  {author} {\bibinfo {author} {\bibfnamefont {F.}~\bibnamefont
  {{Le~Kien}}}, \bibinfo {author} {\bibfnamefont {J.~Q.}\ \bibnamefont
  {Liang}}, \bibinfo {author} {\bibfnamefont {K.}~\bibnamefont {Hakuta}}, \
  and\ \bibinfo {author} {\bibfnamefont {V.~I.}\ \bibnamefont {Balykin}},\
  }\bibfield  {title} {\enquote {\bibinfo {title} {Field intensity
  distributions and polarization orientations in a vacuum-clad
  subwavelength-diameter optical fiber},}\ }\href@noop {} {\bibfield  {journal}
  {\bibinfo  {journal} {Opt. Commun.}\ }\textbf {\bibinfo {volume} {242}},\
  \bibinfo {pages} {445--455} (\bibinfo {year} {2004})}\BibitemShut {NoStop}%
\bibitem [{\citenamefont {Knight}\ \emph {et~al.}(1997)\citenamefont {Knight},
  \citenamefont {Cheung}, \citenamefont {Jacques},\ and\ \citenamefont
  {Birks}}]{knight_ol_1997}%
  \BibitemOpen
  \bibfield  {author} {\bibinfo {author} {\bibfnamefont {J.~C.}\ \bibnamefont
  {Knight}}, \bibinfo {author} {\bibfnamefont {G.}~\bibnamefont {Cheung}},
  \bibinfo {author} {\bibfnamefont {F.}~\bibnamefont {Jacques}}, \ and\
  \bibinfo {author} {\bibfnamefont {T.~A.}\ \bibnamefont {Birks}},\ }\bibfield
  {title} {\enquote {\bibinfo {title} {Phase-matched excitation of
  whispering-gallery-mode resonances by a fiber taper},}\ }\href@noop {}
  {\bibfield  {journal} {\bibinfo  {journal} {Opt. Lett.}\ }\textbf {\bibinfo
  {volume} {22}},\ \bibinfo {pages} {1129--1131} (\bibinfo {year}
  {1997})}\BibitemShut {NoStop}%
\bibitem [{\citenamefont {Wang}\ \emph {et~al.}(2007)\citenamefont {Wang},
  \citenamefont {Pan},\ and\ \citenamefont {Tong}}]{wang_oc_2007}%
  \BibitemOpen
  \bibfield  {author} {\bibinfo {author} {\bibfnamefont {S.}~\bibnamefont
  {Wang}}, \bibinfo {author} {\bibfnamefont {X.}~\bibnamefont {Pan}}, \ and\
  \bibinfo {author} {\bibfnamefont {L.}~\bibnamefont {Tong}},\ }\bibfield
  {title} {\enquote {\bibinfo {title} {Modeling of nanoparticle-induced
  rayleigh-gans scattering for nanofiber optical sensing},}\ }\href@noop {}
  {\bibfield  {journal} {\bibinfo  {journal} {Opt. Commun.}\ }\textbf {\bibinfo
  {volume} {276}},\ \bibinfo {pages} {293--297} (\bibinfo {year}
  {2007})}\BibitemShut {NoStop}%
\bibitem [{\citenamefont {Nieddu}\ \emph {et~al.}(2016)\citenamefont {Nieddu},
  \citenamefont {Gokhroo},\ and\ \citenamefont {{Nic
  Chormaic}}}]{nieddu_jo_2016}%
  \BibitemOpen
  \bibfield  {author} {\bibinfo {author} {\bibfnamefont {T.}~\bibnamefont
  {Nieddu}}, \bibinfo {author} {\bibfnamefont {V.}~\bibnamefont {Gokhroo}}, \
  and\ \bibinfo {author} {\bibfnamefont {S.}~\bibnamefont {{Nic Chormaic}}},\
  }\bibfield  {title} {\enquote {\bibinfo {title} {Optical nanofibres and
  neutral atoms},}\ }\href@noop {} {\bibfield  {journal} {\bibinfo  {journal}
  {J. Opt.}\ }\textbf {\bibinfo {volume} {18}},\ \bibinfo {pages} {053001}
  (\bibinfo {year} {2016})}\BibitemShut {NoStop}%
\bibitem [{\citenamefont {Sadgrove}\ \emph {et~al.}(2016)\citenamefont
  {Sadgrove}, \citenamefont {Wimberger},\ and\ \citenamefont {{Nic
  Chormaic}}}]{sadgrove_sr_2016}%
  \BibitemOpen
  \bibfield  {author} {\bibinfo {author} {\bibfnamefont {M.}~\bibnamefont
  {Sadgrove}}, \bibinfo {author} {\bibfnamefont {S.}~\bibnamefont {Wimberger}},
  \ and\ \bibinfo {author} {\bibfnamefont {S.}~\bibnamefont {{Nic Chormaic}}},\
  }\bibfield  {title} {\enquote {\bibinfo {title} {Quantum coherent tractor
  beam effect for atoms trapped near a nanowaveguide},}\ }\href@noop {}
  {\bibfield  {journal} {\bibinfo  {journal} {Sci. Rep.}\ }\textbf {\bibinfo
  {volume} {6}},\ \bibinfo {pages} {28905} (\bibinfo {year}
  {2016})}\BibitemShut {NoStop}%
\bibitem [{\citenamefont {Ward}\ \emph {et~al.}(2014)\citenamefont {Ward},
  \citenamefont {Maimaiti}, \citenamefont {Le},\ and\ \citenamefont {{Nic
  Chormaic}}}]{ward_rsi_2014}%
  \BibitemOpen
  \bibfield  {author} {\bibinfo {author} {\bibfnamefont {J.~M.}\ \bibnamefont
  {Ward}}, \bibinfo {author} {\bibfnamefont {A.}~\bibnamefont {Maimaiti}},
  \bibinfo {author} {\bibfnamefont {Vu~H.}\ \bibnamefont {Le}}, \ and\ \bibinfo
  {author} {\bibfnamefont {S.}~\bibnamefont {{Nic Chormaic}}},\ }\bibfield
  {title} {\enquote {\bibinfo {title} {Contributed review: Optical micro- and
  nanofiber pulling rig},}\ }\href@noop {} {\bibfield  {journal} {\bibinfo
  {journal} {Rev. Sci. Instrum.}\ }\textbf {\bibinfo {volume} {85}},\ \bibinfo
  {pages} {111501} (\bibinfo {year} {2014})}\BibitemShut {NoStop}%
\bibitem [{\citenamefont {Love}\ \emph {et~al.}(1991)\citenamefont {Love},
  \citenamefont {Henry}, \citenamefont {Stewart}, \citenamefont {Black},
  \citenamefont {Lacroix},\ and\ \citenamefont {Gonthier}}]{love_ieee_1991}%
  \BibitemOpen
  \bibfield  {author} {\bibinfo {author} {\bibfnamefont {J.~D.}\ \bibnamefont
  {Love}}, \bibinfo {author} {\bibfnamefont {W.~M.}\ \bibnamefont {Henry}},
  \bibinfo {author} {\bibfnamefont {W.~J.}\ \bibnamefont {Stewart}}, \bibinfo
  {author} {\bibfnamefont {R.~J.}\ \bibnamefont {Black}}, \bibinfo {author}
  {\bibfnamefont {S.}~\bibnamefont {Lacroix}}, \ and\ \bibinfo {author}
  {\bibfnamefont {F.}~\bibnamefont {Gonthier}},\ }\bibfield  {title} {\enquote
  {\bibinfo {title} {Tapered single-mode fibres and devices, part~1:
  Adiabaticity criteria},}\ }\href@noop {} {\bibfield  {journal} {\bibinfo
  {journal} {IEE Proceedings J - Optoelectronics}\ }\textbf {\bibinfo {volume}
  {138}},\ \bibinfo {pages} {343--354} (\bibinfo {year} {1991})}\BibitemShut
  {NoStop}%
\bibitem [{\citenamefont {Jung}\ \emph {et~al.}(2008)\citenamefont {Jung},
  \citenamefont {Brambilla},\ and\ \citenamefont {Richardson}}]{jung_oe_2008}%
  \BibitemOpen
  \bibfield  {author} {\bibinfo {author} {\bibfnamefont {Y.}~\bibnamefont
  {Jung}}, \bibinfo {author} {\bibfnamefont {G.}~\bibnamefont {Brambilla}}, \
  and\ \bibinfo {author} {\bibfnamefont {D.~J.}\ \bibnamefont {Richardson}},\
  }\bibfield  {title} {\enquote {\bibinfo {title} {Broadband single-mode
  operation of standard optical fibers by using a sub-wavelength optical wire
  filter},}\ }\href@noop {} {\bibfield  {journal} {\bibinfo  {journal} {Opt.
  Express}\ }\textbf {\bibinfo {volume} {16}},\ \bibinfo {pages} {14661--14667}
  (\bibinfo {year} {2008})}\BibitemShut {NoStop}%
\bibitem [{\citenamefont {Sakurai}(1994)}]{sakurai1994quantum}%
  \BibitemOpen
  \bibfield  {author} {\bibinfo {author} {\bibfnamefont {J.~J.}\ \bibnamefont
  {Sakurai}},\ }\href@noop {} {\emph {\bibinfo {title} {Quantum mechanics}}}\
  (\bibinfo  {publisher} {Addison-Wesley Reading},\ \bibinfo {year}
  {1994})\BibitemShut {NoStop}%
\bibitem [{\citenamefont {Chipman}(2010)}]{optics_handbook}%
  \BibitemOpen
  \bibfield  {author} {\bibinfo {author} {\bibfnamefont {R.~A.}\ \bibnamefont
  {Chipman}},\ }\href@noop {} {\emph {\bibinfo {title} {Handbook of optics:
  Ch.~22 Polarimetry}}}\ (\bibinfo  {publisher} {McGraw-Hill},\ \bibinfo {year}
  {2010})\BibitemShut {NoStop}%
\bibitem [{\citenamefont {Madsen}\ \emph {et~al.}(2016)\citenamefont {Madsen},
  \citenamefont {Baker}, \citenamefont {Rubinsztein-Dunlop},\ and\
  \citenamefont {Bowen}}]{madsen_nl_2016}%
  \BibitemOpen
  \bibfield  {author} {\bibinfo {author} {\bibfnamefont {L.~S.}\ \bibnamefont
  {Madsen}}, \bibinfo {author} {\bibfnamefont {C.}~\bibnamefont {Baker}},
  \bibinfo {author} {\bibfnamefont {H.}~\bibnamefont {Rubinsztein-Dunlop}}, \
  and\ \bibinfo {author} {\bibfnamefont {W.~P.}\ \bibnamefont {Bowen}},\
  }\bibfield  {title} {\enquote {\bibinfo {title} {Nondestructive profilometry
  of optical nanofibers},}\ }\href@noop {} {\bibfield  {journal} {\bibinfo
  {journal} {Nano Lett.}\ }\textbf {\bibinfo {volume} {16}},\ \bibinfo {pages}
  {7333--7337} (\bibinfo {year} {2016})}\BibitemShut {NoStop}%
\bibitem [{\citenamefont {Fatemi}\ \emph {et~al.}(2017)\citenamefont {Fatemi},
  \citenamefont {Hoffman}, \citenamefont {Solano}, \citenamefont {Fenton},
  \citenamefont {Beadie}, \citenamefont {Rolston},\ and\ \citenamefont
  {Orozco}}]{fatemi_o_2017}%
  \BibitemOpen
  \bibfield  {author} {\bibinfo {author} {\bibfnamefont {F.~K.}\ \bibnamefont
  {Fatemi}}, \bibinfo {author} {\bibfnamefont {J.~E.}\ \bibnamefont {Hoffman}},
  \bibinfo {author} {\bibfnamefont {P.}~\bibnamefont {Solano}}, \bibinfo
  {author} {\bibfnamefont {E.~F.}\ \bibnamefont {Fenton}}, \bibinfo {author}
  {\bibfnamefont {G.}~\bibnamefont {Beadie}}, \bibinfo {author} {\bibfnamefont
  {S.~L.}\ \bibnamefont {Rolston}}, \ and\ \bibinfo {author} {\bibfnamefont
  {L.~A.}\ \bibnamefont {Orozco}},\ }\bibfield  {title} {\enquote {\bibinfo
  {title} {Modal interference in optical nanofibers for sub-angstrom radius
  sensitivity},}\ }\href@noop {} {\bibfield  {journal} {\bibinfo  {journal}
  {Optica}\ }\textbf {\bibinfo {volume} {4}},\ \bibinfo {pages} {157--162}
  (\bibinfo {year} {2017})}\BibitemShut {NoStop}%
\bibitem [{\citenamefont {Picardi}\ \emph {et~al.}(2018)\citenamefont
  {Picardi}, \citenamefont {Zayats},\ and\ \citenamefont {Rodr\'iguez-Fortu{\~
  n}o}}]{picardi_prl_2018}%
  \BibitemOpen
  \bibfield  {author} {\bibinfo {author} {\bibfnamefont {M.~F.}\ \bibnamefont
  {Picardi}}, \bibinfo {author} {\bibfnamefont {A.~V.}\ \bibnamefont {Zayats}},
  \ and\ \bibinfo {author} {\bibfnamefont {F.~J.}\ \bibnamefont
  {Rodr\'iguez-Fortu{\~ n}o}},\ }\bibfield  {title} {\enquote {\bibinfo {title}
  {Janus and huygens dipoles: near-field directionality beyond spin-momentum
  locking},}\ }\href@noop {} {\bibfield  {journal} {\bibinfo  {journal} {Phys.
  Rev. Lett.}\ }\textbf {\bibinfo {volume} {120}},\ \bibinfo {pages} {117402}
  (\bibinfo {year} {2018})}\BibitemShut {NoStop}%
\bibitem [{\citenamefont {Petersen}\ \emph {et~al.}(2014)\citenamefont
  {Petersen}, \citenamefont {Volz},\ and\ \citenamefont
  {Rauschenbeutel}}]{petersen_s_2014}%
  \BibitemOpen
  \bibfield  {author} {\bibinfo {author} {\bibfnamefont {J.}~\bibnamefont
  {Petersen}}, \bibinfo {author} {\bibfnamefont {J.}~\bibnamefont {Volz}}, \
  and\ \bibinfo {author} {\bibfnamefont {A.}~\bibnamefont {Rauschenbeutel}},\
  }\bibfield  {title} {\enquote {\bibinfo {title} {Chiral nanophotonic
  waveguide interface based on spin-orbit interaction of light},}\ }\href@noop
  {} {\bibfield  {journal} {\bibinfo  {journal} {Science}\ }\textbf {\bibinfo
  {volume} {346}},\ \bibinfo {pages} {67--71} (\bibinfo {year}
  {2014})}\BibitemShut {NoStop}%
\bibitem [{\citenamefont {Sadgrove}\ \emph {et~al.}(2017)\citenamefont
  {Sadgrove}, \citenamefont {Sugawara}, \citenamefont {Mitsumori},\ and\
  \citenamefont {Edamatsu}}]{sadgrove_sr_2017}%
  \BibitemOpen
  \bibfield  {author} {\bibinfo {author} {\bibfnamefont {M.}~\bibnamefont
  {Sadgrove}}, \bibinfo {author} {\bibfnamefont {M.}~\bibnamefont {Sugawara}},
  \bibinfo {author} {\bibfnamefont {Y.}~\bibnamefont {Mitsumori}}, \ and\
  \bibinfo {author} {\bibfnamefont {K.}~\bibnamefont {Edamatsu}},\ }\bibfield
  {title} {\enquote {\bibinfo {title} {Polarization response and scaling law of
  chirality for a nanofibre optical interface},}\ }\href@noop {} {\bibfield
  {journal} {\bibinfo  {journal} {Sci. Rep.}\ }\textbf {\bibinfo {volume}
  {7}},\ \bibinfo {pages} {1--9} (\bibinfo {year} {2017})}\BibitemShut
  {NoStop}%
\bibitem [{\citenamefont {Bliokh}\ \emph
  {et~al.}(2015{\natexlab{a}})\citenamefont {Bliokh}, \citenamefont
  {Rodr\'iguez-Fortu{\~n}o}, \citenamefont {Nori},\ and\ \citenamefont
  {Zayats}}]{bliokh_np_2015}%
  \BibitemOpen
  \bibfield  {author} {\bibinfo {author} {\bibfnamefont {K.~Y.}\ \bibnamefont
  {Bliokh}}, \bibinfo {author} {\bibfnamefont {F.~J.}\ \bibnamefont
  {Rodr\'iguez-Fortu{\~n}o}}, \bibinfo {author} {\bibfnamefont
  {F.}~\bibnamefont {Nori}}, \ and\ \bibinfo {author} {\bibfnamefont {A.~V.}\
  \bibnamefont {Zayats}},\ }\bibfield  {title} {\enquote {\bibinfo {title}
  {Spin-orbit interactions of light},}\ }\href@noop {} {\bibfield  {journal}
  {\bibinfo  {journal} {Nat. Photonics}\ }\textbf {\bibinfo {volume} {9}},\
  \bibinfo {pages} {796--808} (\bibinfo {year}
  {2015}{\natexlab{a}})}\BibitemShut {NoStop}%
\bibitem [{\citenamefont {{Van Mechelen and J.
  Zubin}}(2016)}]{vanMechelen_o_2016}%
  \BibitemOpen
  \bibfield  {author} {\bibinfo {author} {\bibfnamefont {T.}~\bibnamefont {{Van
  Mechelen and J. Zubin}}},\ }\bibfield  {title} {\enquote {\bibinfo {title}
  {Universal spin-momentum locking of evanescent waves},}\ }\href@noop {}
  {\bibfield  {journal} {\bibinfo  {journal} {Optica}\ }\textbf {\bibinfo
  {volume} {3}},\ \bibinfo {pages} {118--126} (\bibinfo {year}
  {2016})}\BibitemShut {NoStop}%
\bibitem [{\citenamefont {O'Connor}\ \emph {et~al.}(2014)\citenamefont
  {O'Connor}, \citenamefont {Ginzburg}, \citenamefont {Rodr\'iguez-Fortu{\~
  n}o}, \citenamefont {Wurtz},\ and\ \citenamefont {Zayats}}]{oconnor_nc_2014}%
  \BibitemOpen
  \bibfield  {author} {\bibinfo {author} {\bibfnamefont {D.}~\bibnamefont
  {O'Connor}}, \bibinfo {author} {\bibfnamefont {P.}~\bibnamefont {Ginzburg}},
  \bibinfo {author} {\bibfnamefont {F.~J.}\ \bibnamefont {Rodr\'iguez-Fortu{\~
  n}o}}, \bibinfo {author} {\bibfnamefont {G.~A.}\ \bibnamefont {Wurtz}}, \
  and\ \bibinfo {author} {\bibfnamefont {A.~V.}\ \bibnamefont {Zayats}},\
  }\bibfield  {title} {\enquote {\bibinfo {title} {Spin–orbit coupling in
  surface plasmon scattering by nanostructures},}\ }\href@noop {} {\bibfield
  {journal} {\bibinfo  {journal} {Nat. Commun.}\ }\textbf {\bibinfo {volume}
  {5}},\ \bibinfo {pages} {5327} (\bibinfo {year} {2014})}\BibitemShut
  {NoStop}%
\bibitem [{\citenamefont {Bliokh}\ \emph
  {et~al.}(2015{\natexlab{b}})\citenamefont {Bliokh}, \citenamefont
  {Smirnova},\ and\ \citenamefont {Nori}}]{bliokh_s_2015}%
  \BibitemOpen
  \bibfield  {author} {\bibinfo {author} {\bibfnamefont {K.~Y.}\ \bibnamefont
  {Bliokh}}, \bibinfo {author} {\bibfnamefont {D.}~\bibnamefont {Smirnova}}, \
  and\ \bibinfo {author} {\bibfnamefont {F.}~\bibnamefont {Nori}},\ }\bibfield
  {title} {\enquote {\bibinfo {title} {Quantum spin hall effect of light},}\
  }\href@noop {} {\bibfield  {journal} {\bibinfo  {journal} {Science}\ }\textbf
  {\bibinfo {volume} {348}},\ \bibinfo {pages} {1448--1451} (\bibinfo {year}
  {2015}{\natexlab{b}})}\BibitemShut {NoStop}%
\bibitem [{Note1()}]{Note1}%
  \BibitemOpen
  \bibinfo {note} {For the ease of presentation, this figure shows experimental
  data collected after the polarization compensation.}\BibitemShut {Stop}%
\bibitem [{\citenamefont {{Snyder and J.~D.~Love}}(1983)}]{snyder_book}%
  \BibitemOpen
  \bibfield  {author} {\bibinfo {author} {\bibfnamefont {A.~W.}\ \bibnamefont
  {{Snyder and J.~D.~Love}}},\ }\href@noop {} {\emph {\bibinfo {title} {Optical
  waveguide theory}}}\ (\bibinfo  {publisher} {Chapman and Hall},\ \bibinfo
  {year} {1983})\BibitemShut {NoStop}%
\end{thebibliography}%
\end{document}